\documentclass[11pt,a4paper,preprintnumbers,nofootinbib]{revtex4}
\usepackage[T1]{fontenc}
\usepackage[utf8]{inputenc}
\usepackage{mathtools}
\usepackage{amsmath}
\usepackage{amsthm}
\usepackage{amsmath}
\usepackage{amssymb}
\usepackage{amstext}
\usepackage[final]{graphicx}
\usepackage{float}
\usepackage{booktabs}
\usepackage{fixltx2e}
\usepackage{subcaption}
\usepackage{siunitx}
\usepackage{tikz}
\usepackage{tikz-feynman}
\usepackage{array}
\usepackage{slashed}
\usepackage[section]{placeins}
\usepackage{tabularx}
\PassOptionsToPackage{hyphens}{url}
\usepackage{hyperref}
\usepackage{makecell}	
\allowdisplaybreaks
\usepackage[normalem]{ulem}

\newcommand{\mytitle}{Missing Energy plus Jet in the  SMEFT}
\usepackage{braket}
\begin{document}

\setlength{\parindent}{0pt}

\title{\mytitle}

\author{Gudrun Hiller}
\email{ghiller@physik.uni-dortmund.de}
\author{Daniel Wendler}
\email{daniel.wendler@tu-dortmund.de}
\affiliation{TU Dortmund University, Department of Physics, Otto-Hahn-Str.4, D-44221 Dortmund, Germany}

\begin{abstract}
We study the production of  dineutrinos  in proton-proton collisions,  with  large missing transverse energy and an energetic jet as the experimental signature.
Recasting a  search from the ATLAS collaboration
we work out  constraints on semileptonic four-fermion operators, gluon and electroweak dipole operators and $Z$-penguins in the SMEFT.
All but the $Z$-penguin operators experience energy-enhancement.
Constraints on gluon dipole operators are the strongest,  probing new physics up to  14 TeV,
and improve over existing ones from collider studies.
Limits on FCNC  four-fermion operators are  competitive  with Drell-Yan production of dileptons, and improve on those  for  tau  final states.
For left-handed $|\Delta s|=|\Delta d|=1$ and right-handed $|\Delta c|=|\Delta u|=1$ transitions these are the best available limits, also considering rare kaon and charm decays.
We estimate  improvements  for the $\SI{3000}{\femto\barn}^{-1}$ High Luminosity Large Hadron Collider.

\end{abstract}

\maketitle

\tableofcontents

\section{Introduction}

The Standard Model Effective Field Theory (SMEFT) \cite{Buchmuller:1985jz,Grzadkowski:2010es} is a framework  to study  new physics from beyond the electroweak scale without
resorting to a model. It allows for a powerful joint interpretation of  data from different experiments, sectors  and energy scales in particular when combined with
the weak effective theory (WET). 
Recent global analyses combine  electroweak precision, top quark observables and Drell-Yan  \cite{Drell:1970wh} production of  charged leptons, 
 with rare $B$-decays \cite{Aoude:2020dwv,Bruggisser:2021duo,Greljo:2022jac,Grunwald:2023nli}. 
 Drell-Yan studies have received attention  because of the energy enhancement 
  of operator insertions, that amplifies their contributions kinematically
 and competes with the suppression of scales \cite{Farina:2016rws}.
The quark flavors available from the colliding  protons also make the Large Hadron Collider (LHC) a flavor factory and Drell-Yan studies a probe of flavor physics.
  In addition, different operators typically  contribute  incoherently in the high energy limit, hence large cancellations are avoided
  \cite{Greljo:2017vvb,Fuentes-Martin:2020lea}.

In this work we extend  this  type of standard model (SM) tests  by considering the 'Drell-Yan' production of dineutrinos in proton-proton collisions, where the experimental signature is  large missing transverse energy (MET) and an energetic jet.
We consider flavorful operators of dimension six, inducing flavor-changing neutral  currents (FCNCs) of quarks.
Generally $\nu \bar \nu$ production is described by the Drell-Yan process,  such as $Z$-exchange in the SM,
however  $q \bar q \rightarrow \nu \bar \nu$ does not generate any visible
transverse momentum and therefore  we consider an energetic jet in addition to transverse missing energy.
While the extra jet implies a suppression with the strong coupling constant, the cross sections  to dineutrinos are enhanced by the  sum over all flavors, including taus, as well as large coupling to the $Z$.
We recast the ATLAS search  \cite{ATLAS:2021kxv} 
based on $\mathcal{L}_{int}= \SI{139}{\femto\barn}^{-1}$.
Furthermore a CMS data set \cite{CMS:2021far} based on $\mathcal{L}_{int}= \SI{101}{\femto\barn}^{-1}$ exist, which was not considered in this work.
Projections are  derived for the High Luminosity Large Hadron Collider (HL-LHC) \cite{Cepeda:2019klc}, assuming naive statistical scaling of the dataset  with integrated luminosity of  $\SI{3000}{\femto\barn}^{-1}$.

The paper is organized as follows:
In Sec.~\ref{sec:framework} we introduce the SMEFT framework. We analyze the missing transverse energy distributions in Sec.~\ref{sec:Emiss},
where we also discuss flavor hierarchies and energy enhancement.
The recast of the ATLAS analysis \cite{ATLAS:2021kxv} is worked out in Sec.~\ref{sec:simulation}.
We present the results of our analysis and compare the bounds to those from other observables in Sec.~\ref{sec:bounds}.
We summarize in Sec.~\ref{sec:sum}.
In App.~\ref{app:WET} we briefly  introduce the  WET  relevant for semleptonic four-fermion and dipole operators.
In App.~\ref{app:pertub} we present a fully analytic framework and derive analytical expressions for the partonic differential cross sections contributing to
$\nu \bar \nu$ plus a hard jet  at leading order in the SM and in SMEFT.

\section{SMEFT framework}

\label{sec:framework}

We present the set-up of the SMEFT analysis in Sec.~\ref{sec:setup}. and discuss  semileptonic four-fermion operators (Sec.~\ref{sec:fourfermion}), gluon dipole operators
(Sec.~\ref{sec:gluon-dipole}) and electroweak dipole and penguin operators (Sec.~\ref{sec:penguin}) which contribute  at tree-level to  $pp \rightarrow \nu \bar \nu + X$.

\subsection{Set-up \label{sec:setup}}

The SMEFT lagrangian is  the one of the SM,  $\mathcal{L}_{\text{SM}}$,  plus an infinite  tower of higher dimensional operators
$Q_x^{(d)}$  of mass dimension $d$  composed out of SM fields, respecting  SM gauge and Poincare invariance, that can be written as
\begin{equation} \nonumber
      \mathcal{L}_{\text{SMEFT}}\,=\, \mathcal{L}_{\text{SM}}\,+ \,\sum_{d \geq 5} \,\sum_x\: \frac{{C_x^{(d)}}}{\Lambda_{NP}^{d-4}}{Q_x^{(d)}}  \, .
\end{equation}
Operators with the same dimension are distinguished by a label $x$.
The $C_x^{(d)}$ are the corresponding Wilson coefficients and
$\Lambda_{NP}$ denotes the scale of NP, which is assumed to be sufficiently separated from  the electroweak scale  set by the vacuum expectation value of the
Higgs field, $v \simeq 246$ GeV.
Operators with lower  dimension are  less suppressed  by powers of the NP scale, and matter more for phenomenology.
Since we do not consider  baryon- and lepton-number violating  processes, which involve odd dimension, we focus on  $d=6$,
and drop the superscript '$(d)$'  from now on.

Dimension $6$ operators contributing  to the process $ p p \rightarrow \nu \bar \nu + X$ are listed in Table \ref{tab:operators}, based on the Warsaw basis \cite{Grzadkowski:2010es}.
They belong to four  categories: semileptonic four-fermion operators, electroweak  (EW)  and gluonic dipole operators  and $Z$-penguin operators,  labelled by $4F, EW, G$ and $ZP$, respectively.
The neutrinos are left-handed and  contained in the lepton doublets, $l$. The quark doublets are denoted by $q$, and up-type (down-type) singlets by $u (d)$.
The field strength tensors of the $SU(3)_C \times SU(2)_L \times U(1)_Y$ bosons are denoted by $G^A_{\mu \nu}, W^I_{\mu \nu}$ and $B_{\mu \nu}$, 
where $T^A=\lambda^A/2 $, $A=1, \ldots , 8$ and $\tau^I/2$, $I=1,2,3$  are the generators of $SU(3)_C$ and $SU(2)_L $, respectively, in the fundamental representation. Here, $\lambda^A$ are the Gell-Mann matrices and $\tau^I$ the Pauli-matrices.
$\varphi$ is the Higgs and $\tilde\varphi= i \tau^2 \varphi$  its conjugate, $ \left(\varphi^\dagger i\overleftrightarrow D_\mu \varphi\right)=i\varphi^\dagger( D_\mu \varphi )-  i( D_\mu \varphi)^\dagger \varphi $ and $\left(\varphi^\dagger i\overleftrightarrow D_\mu^I \varphi\right)=i\varphi^\dagger \tau^I ( D_\mu \varphi )-  i( D_\mu \varphi)^\dagger \tau^I \varphi $
and $D_\mu$ is the {covariant derivative.
Quark (lepton) flavor indices are denoted by $i,j$ ($k,l$).
Note that additional operators, which modify the $\nu \bar \nu Z$ vertex, are not considered in this work.
\begin{table}
 \centering
    \begin{tabular}{@{} |ll||ll| @{}}
      \toprule 
      \multicolumn{1}{|l|}{$Q^{(1)}_{lq}$}        &\multicolumn{1}{|l|}{$\left(\bar l_k \gamma_{\mu}  l_l \right)\left( \bar q_i \gamma^{\mu}  q_j \right) $}                    &          \multicolumn{1}{|l|}{$Q_{lu}$}                        & \multicolumn{1}{|l|}{$( \bar l_k \gamma_{\mu}l_l   ) (\bar u_i \gamma^{\mu} u_j)$}            \\
      \multicolumn{1}{|l|}{$Q^{(3)}_{lq}$}        &\multicolumn{1}{|l|}{$\left(\bar l_k \gamma_{\mu} \tau^I l_l \right)\left( \bar q_i \gamma^{\mu} \tau^I q_j \right) $ }       &          \multicolumn{1}{|l|}{$Q_{ld}$}                        & \multicolumn{1}{|l|}{$( \bar l_k \gamma_{\mu}l_l   ) (\bar d_i \gamma^{\mu} d_j)$}            \\ \midrule
      \multicolumn{1}{|l|}{$Q_{uG}$}              & \multicolumn{1}{|l|}{$(\bar q_i \sigma^{\mu \nu} T^A u_j  \tilde \varphi G^A_{\mu \nu}     )$}                               &           \multicolumn{1}{|l|}{$Q_{\varphi u }  $}              &    \multicolumn{1}{|l|}{$ \left( \varphi^{\dagger} i \stackrel{\leftrightarrow}{D_{\mu}} \varphi \right)  \left( \bar u_i \gamma^{\mu} u_j \right) $}                       \\
      \multicolumn{1}{|l|}{$Q_{dG}$}              & \multicolumn{1}{|l|}{$(\bar q_i \sigma^{\mu \nu} T^A d_j   \varphi G^A_{\mu \nu}     )$}                                     &           \multicolumn{1}{|l|}{$Q_{\varphi d }  $}              &   \multicolumn{1}{|l|}{$ \left( \varphi^{\dagger} i \stackrel{\leftrightarrow}{D_{\mu}} \varphi \right)  \left( \bar d_i \gamma^{\mu} d_j \right) $}                                                                                                                                            \\
      \multicolumn{1}{|l|}{$Q_{uW}$}              & \multicolumn{1}{|l|}{$(\bar q_i \sigma^{\mu \nu}  \tau^I u_j \tilde \varphi W^I_{\mu \nu}     )$}                            &           \multicolumn{1}{|l|}{$Q^{(1)}_{\varphi q }  $}        &   \multicolumn{1}{|l|}{$ \left( \varphi^{\dagger} i \stackrel{\leftrightarrow}{D_{\mu}} \varphi \right)  \left( \bar q_i \gamma^{\mu} q_j \right) $}                                                                                                                                                    \\
      \multicolumn{1}{|l|}{$Q_{dW}$}              &  \multicolumn{1}{|l|}{$(\bar q_i \sigma^{\mu \nu}  \tau^I d_j  \varphi W^I_{\mu \nu}     )$}                                 &           \multicolumn{1}{|l|}{$Q^{(3)}_{\varphi q }  $}        &    \multicolumn{1}{|l|}{$ \left( \varphi^{\dagger} i \stackrel{\leftrightarrow}{D_{\mu}^I} \varphi \right)\left( \bar q_i \tau^I \gamma^{\mu} q_j \right) $}                             \\ 
      \multicolumn{1}{|l|}{$Q_{uB}$}              &  \multicolumn{1}{|l|}{$(\bar q_i \sigma^{\mu \nu}   u_j \tilde \varphi B_{\mu \nu}     )$}                                   &              \multicolumn{1}{|l|}{}                                                     &        \multicolumn{1}{|l|}{}                                           \\
      \multicolumn{1}{|l|}{$Q_{dB}$}              &   \multicolumn{1}{|l|}{$(\bar q_i \sigma^{\mu \nu}   d_j  \varphi B_{\mu \nu}     )$}                                  &              \multicolumn{1}{|l|}{}                                                     &        \multicolumn{1}{|l|}{}                             \\ \bottomrule
\end{tabular}
  \caption{Dimension-6 SMEFT operators contributing to the process $ pp \rightarrow \nu \bar \nu + X$,
  four-fermion operators (upper two rows), gluon  and electroweak  dipoles (bottom left), and electroweak penguins (bottom right).  See text for details. }
  \label{tab:operators}
\end{table} 

The operators in Table \ref{tab:operators} are given in the flavor basis of the fermions. 
Quark mass and flavor bases are related by unitary transformations, whose net effect in the SM concerns the  quark doublets and  is contained in the 
Cabibbo-Kobayashi-Maskawa (CKM) mixing matrix  $V$.
The related effect  from the basis change for the doublet leptons cancels in the dineutrino observables  due to unitarity of the 
Pontecorvo-Maki-Nakagawa-Sakata (PMNS)-mixing matrix after summing over all neutrino flavors \cite{Bause:2020auq}.
Since the rotation of  quark singlets is unphysical in the SM, we absorb it into the WCs of $Q_{lu}, Q_{ld}, Q_{\varphi u}, Q_{\varphi d}$  \cite{Aebischer:2015fzz}.
For the dipole operators into neutral currents in addition we assume when working out bounds for up-sector (down-sector) FCNCs that the quarks are in the up-mass basis  (down mass basis), and corresponding bounds are understood in this basis.
Operators with doublet quark currents, $Q_{lq}^{(1,3)}, Q_{\varphi q}^{(1,3)}$ are more complicated, as
$SU(2)_L$ dictates  that switching on up-sector currents unavoidably imply down-sector ones, and vice versa
 \cite{Bissmann:2020mfi}. We discuss this  further in Sec.~\ref{sec:fourfermion} on four-fermion operators and in Sec.~\ref{sec:penguin} for the $Z$-penguins.

Generically, within $d=6$ SMEFT a cross section  $\sigma$ can be parametrized as 
\begin{equation} \label{eq:sig}
  \sigma = \sigma^{SM} + \sum_x \frac{C_x}{\Lambda_{NP}^2} \sigma^{int}_{x} + \sum_{x \geq y} \frac{C_x C_y^{*}}{\Lambda_{NP}^4} \sigma^{NP}_{xy},
\end{equation}
where  the first, second and third term corresponds to the SM contribution, a SM-NP interference term and the pure NP contribution, respectively, and 
the Wilson coefficients  $C_x$ and the NP scale $\Lambda_{NP}$ have been  factored out.
In this work we focus on FCNC operator insertions, for which  $\sigma^{int}_{xy}$  vanishes in the limit that SM-FCNCs, which are loop-, Glashow-Iliopoulos-Maiani~(GIM)- and CKM-suppressed, are neglected.
This also avoids contributions to (\ref{eq:sig}) from FCNC $d=8$ operators  at $\mathcal{O}(1/\Lambda_{NP}^4)$ via SM-NP interference.

As just argued,  we obtain limits on $d=6$ SMEFT coefficients at leading order which is  $\mathcal{O}(1/\Lambda_{NP}^4)$.
This is not necessarily a too strong of a suppression to reach  higher energies due to  the  energy enhancement of operators  $\sigma^{NP}_{xy} \sim E^2$,  where 
$E$ stands for a, or a set of   kinematic variables such as the parton's energy,
rather than 
the naive scale suppression $\sigma^{NP}_{xy} \sim v^2$, e.g.~\cite{Farina:2016rws}. Given the energy reach of the LHC together with high luminosities expected for the HL-LHC,  this allows to probe NP indirectly in high $p_T$-tails  up to the  few $O(10)$-TeV region \cite{Cepeda:2019klc}.  We observe and  discuss  energy-enhanced  differential cross sections for 
 $pp \to \text{MET} +j$ in Sec.~\ref{sec:shapes}.

\subsection{Semileptonic four-fermion operators}
\label{sec:fourfermion}

Here we consider semileptonic four-fermion operators in the SMEFT framework that contribute to $p p \rightarrow \nu \bar \nu + X$. The corresponding  Lagrangian reads  
\begin{equation*}
    \label{eqn:lagrange_dim6}
    \begin{aligned}
      \mathcal{L}^{6}_{\text{SMEFT}}  \supset& \quad \frac{ C^{(1)}_{lq,kl ij }}{\Lambda_{NP}^2}\left(\bar l_k \gamma_{\mu}  l_l \right)\left( \bar q_i \gamma^{\mu}  q_j \right) +   \frac{C^{(3)}_{lq,kl ij }}{\Lambda_{NP}^2}\left(\bar l_k \gamma_{\mu} \tau^{I}  l_l \right)\left( \bar q_i \gamma^{\mu} \tau^{I}  q_j \right)& \\
                  & + \frac{C_{lu,kl ij }}{\Lambda_{NP}^2}\left(\bar l_k \gamma_{\mu}  l_l \right)\left( \bar u_i \gamma^{\mu}  u_j \right) +\frac{C_{ld,kl ij }}{\Lambda_{NP}^2}\left(\bar l_k \gamma_{\mu}  l_l \right)\left( \bar d_i \gamma^{\mu}  d_j \right) \, .
    \end{aligned}
  \end{equation*}
Feynman diagrams for the processes $ q_i g \rightarrow q_j \nu_k \bar \nu_l$ and $q_i \bar q_j \rightarrow g \nu_k \bar \nu_l$, with insertions of semileptonic four-fermion operators, are depicted in Fig.~\ref{fig:parton_feyn}. 
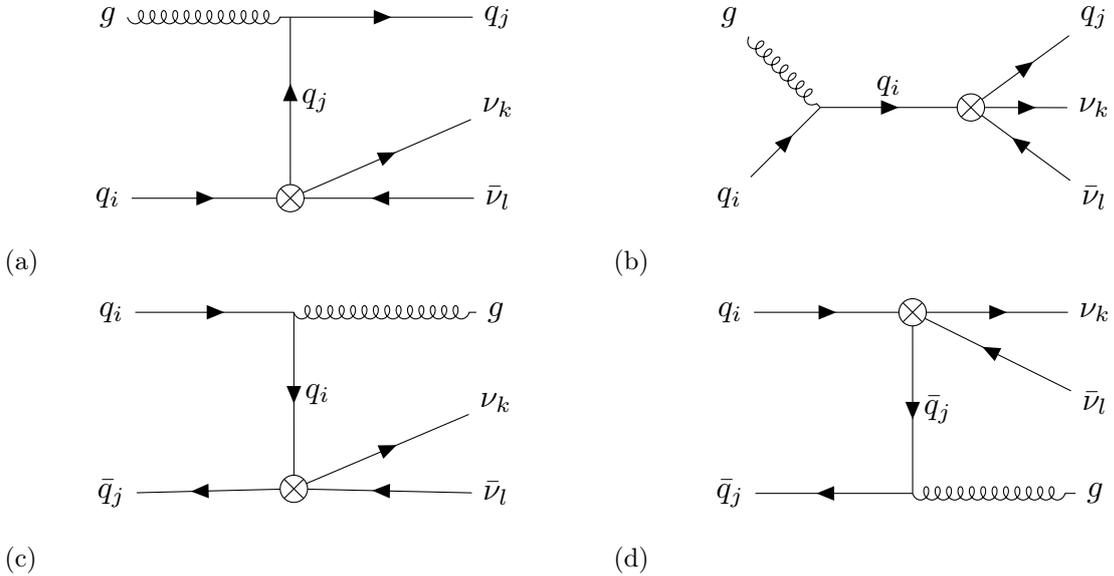
\begin{figure}[b]
    \centering
    \begin{subfigure}{0.48 \textwidth}
      \centering
      \begin{tikzpicture}[scale=1.2,transform shape]
        \begin{feynman}
        \vertex (i1) {\(g\)};
        \vertex [right = 2cm of i1] (a);
        \vertex [crossed dot,below = 1.85cm of a] (b){};
        \vertex [right = 2cm of b ] (c);
        \vertex [below = 2cm of i1] (i2) {\(q_i\)};
        \vertex [right = 2cm of a] (f1){\(q_j\)};
        \vertex [below = 1cm of f1] (f2){\(\nu_k\)};
        \vertex [below = 2cm of f1 ] (f3) {\( \bar \nu_l\)}; 
  
        \diagram* {
        (i1) -- [gluon](a),
        (i2) -- [fermion](b),
        (a) -- [anti fermion,edge label = \(q_j\)](b),
        (a) -- [ fermion] (f1),
        (b) -- [fermion] (f2),
        (b) -- [anti fermion] (f3),
        };
        \end{feynman}
        
    \end{tikzpicture}
    \caption{}
    \label{fig:qg_t_channel}
    \end{subfigure}
    \begin{subfigure}{0.48 \textwidth}
      \centering
      \begin{tikzpicture}[scale=1.2,transform shape]
        \begin{feynman}
        \vertex (i1) {\(g\)};
        \vertex [below right = 1.41cm  of i1] (a);
        \vertex [crossed dot,right =1.5cm of a] (b){};
        \vertex [ right = 3cm of i2](c);
        \vertex [below  =2cm of i1] (i2) {\(q_i\)};
        \vertex [ right =4cm of i1] (f1){\(q_j\)};
        \vertex [below = 1cm of f1] (f2){\(\nu_k\)};
        \vertex [below = 2cm of f1] (f3) {\( \bar \nu_l\)}; 
  
        \diagram* {
            (i1) -- [gluon](a),
            (i2) -- [fermion](a),
            (a) -- [fermion,edge label = \(q_i\)](b),
            (b) -- [ fermion] (f1),
            (b) -- [fermion] (f2),
            (b) -- [anti fermion] (f3),
            };
        \end{feynman}
        
    \end{tikzpicture}
    \caption{}
    \label{fig:qg_s_channel}
    \end{subfigure}
    \begin{subfigure}{0.48 \textwidth}
      \centering
      \begin{tikzpicture}[scale=1.2,transform shape]
        \begin{feynman}
          \vertex (i1) {\( q_i\)};
          
          \vertex [ right= 2cm of i1](a) ;
          \vertex [crossed dot, below= 1.8cm of a](b){};
          \vertex [right = 2cm of b] (c);
          \vertex [below  = 2cm of i1](i2) {\( \bar q_j\)};
          \vertex [ right= 2cm of a](f1){\(g\)};
          \vertex [ below = 1cm of f1](f3){\( \nu_k\)};
          \vertex [below  = 1cm of f3](f2){\(\bar\nu_l\)};
          \diagram* {
            (i1) -- [ fermion](a),
            (i2) -- [anti fermion](b),
            (a) -- [ fermion,edge label = \(q_i\)](b),
            (b) -- [anti fermion] (f2),
            (b) -- [fermion] (f3),
            (a) -- [gluon] (f1),
            };
        \end{feynman}
      \end{tikzpicture}
      \caption{}
      \label{fig:qq_t_channel}
    \end{subfigure}
    \begin{subfigure}{0.48 \textwidth}
      \centering
      \begin{tikzpicture}[scale=1.2,transform shape]
        \begin{feynman}
          \vertex (i1) {\( q_i\)};
          
          \vertex [crossed dot, right= 2cm of i1](a) {};
          \vertex [ below= 2cm of a](b);
          \vertex [right = 2cm of b] (c);
          \vertex [below  = 2cm of i1](i2) {\(\bar q_j\)};

          \vertex [ right = 2       of a](f3){\( \nu_k\)};
          \vertex [below  = 1cm of f3](f2){\(\bar\nu_l\)};
          \vertex [ below = 2cm  of f3](f1){\(g\)};
          \diagram* {
            (i1) -- [ fermion](a),
            (i2) -- [ anti fermion](b),
            (a) -- [ fermion,edge label =\(\bar q_j\) ](b),
            (a) -- [anti fermion] (f2),
            (a) -- [fermion] (f3),
            (b) -- [gluon] (f1),
            };
        \end{feynman}
      \end{tikzpicture}
      \caption{}
      \label{fig:qq_u_channel}
    \end{subfigure}
    \caption{Parton level diagrams contributing to $p p \rightarrow \nu \bar \nu + X$ with  insertions of semileptonic four-fermion operators. The diagrams (a) and (b) show the contributions corresponding to  $q_ig \rightarrow q_j \nu_k \bar \nu_l$, while (c) and (d) the ones for  $q_i \bar q_j \rightarrow g \nu_k \bar \nu_l$.}
    \label{fig:parton_feyn}
  \end{figure}
The partonic cross sections are proportional to the square of the effective WC
\begin{equation}
    \label{eqn:WC_eff}
    C^{4F}_{ij} =\sqrt{ \sum_{k,l} |C^{\pm}_{kl ij}|^2 + |C_{l u/d ,kl ij}|^2},
\end{equation}
where
\begin{equation*}
    C^{\pm}_{ijkl} = C^{(1)}_{lq,kl ij} \pm C^{(3)}_{lq,kl ij}.
\end{equation*} 
The sum runs over all lepton flavors, and both right- and left-chiral WCs contribute equally since all masses are neglected.
Note that the $\pm $ and $u/d$ depends on the quark sector.
Explicity, $ C^{+}_{ijkl} $ contributes to up-type quarks with dineutrinos, as well as down-type quarks with  charged dileptons, whereas 
$ C^{-}_{ijkl} $ induces 
the $SU(2)_L$-flipped
 transitions, that is, down-type quarks with dineutrinos, plus up-type quarks with  charged dileptons.  It is evident that 
 Drell-Yan production of dineutrinos, which we are exploring in this work, should  be analyzed together with charged-lepton data. We hope to come back to this in the future.
 
Limits  on the effective WC \eqref{eqn:WC_eff} can  be used to derive constraints on individual WCs with specific chirality and flavor.
Within more specific models, one  can also derive stronger bounds.
For instance, assuming equal  left- and right-chiral WCs, $C^{\pm}_{kl ij}= C_{l u/d ,kl ij}$,  one obtains
\begin{equation*}
  \begin{aligned}
  \left( C^{4F}_{ij} \right)^2 
               &= 2 \sum_{k,l} |C^{\pm}_{kl ij}|^2,
    \end{aligned}
\end{equation*}
where each individual flavor combination $ij$  of the WC  is constrained stronger by a factor of $\frac{1}{\sqrt{2}}$ than in the general case  (\refeq{eqn:WC_eff}).
Another application are flavor symmetries, such as a  $U(3)_l \times U(3)_e $ lepton flavor symmetry, where the WCs are lepton flavor universal as
$C^{\pm}_{kl ij} =\delta_{kl} C^{\pm}_{ij}$ and $C_{l u/d ,kl ij} =\delta_{kl} C_{l u/d ,ij}$ and the sums  over the neutrino flavors collapse to
\begin{equation*}
  \begin{aligned}
  \left( C^{4F}_{ij} \right)^2 
               &=   3\left(|C^{\pm}_{ ij}|^2 + |C_{l u/d ij}|^2\right) \, .
              \end{aligned}
            \end{equation*}
The left- and right-chiral WCs are each constrained stronger by  a factor of $\frac{1}{\sqrt{3}}$  than in (\refeq{eqn:WC_eff}).

\subsection{Gluon dipole operators}
\label{sec:gluon-dipole}

The $q\bar qg$ vertex gets modified by the insertion of a gluonic dipole operator.
The effective  Lagrangian reads
\begin{equation*}
  \label{eqn:lagr_dipole}
  \mathcal{L}^{6}_{\text{SMEFT}} \supset \frac{C_{uG,ij} }{\Lambda_{NP}^2}  \left(\bar q_i \sigma^{\mu \nu} T^A u_j \right) \tilde \varphi G^{A}_{\mu \nu} +\frac{C_{dG,ij} }{\Lambda_{NP}^2}  \left(\bar q_i \sigma^{\mu \nu} T^A d_j \right)  \varphi G^{A}_{\mu \nu}  
\end{equation*}
 and the diagrams with the corresponding insertions are depicted in Fig.~\ref{fig:Dipole}.
\begin{figure}[b]
  \centering
  \begin{subfigure}{0.48 \textwidth}
    \centering
    \begin{tikzpicture}[scale=1.2,transform shape]
      \begin{feynman}
      \vertex (i1) {\(g\)};
      \vertex [crossed dot,right = 2cm of i1] (a){};
      \vertex [below = 2cm of a] (b);
      \vertex [right = 2cm of b ] (c);
      \vertex [below = 2cm of i1] (i2) {\(q_i\)};
      \vertex [right = 1cm of b] (d);
      \vertex [right = 2cm of a] (f1){\(q_j\)};
      \vertex [below = 1cm of f1] (f2){\(\nu\)};
      \vertex [below = 2cm of f1 ] (f3) {\( \bar \nu\)}; 
      \diagram* {
      (i1) -- [gluon](a),
      (i2) -- [fermion](b),
      (a) -- [anti fermion,edge label = \(q_i\)](b),
      (a) -- [ fermion] (f1),
      (b) -- [ boson,edge label = \(Z\)] (d),
      (d) -- [fermion] (f2),
      (d) -- [anti fermion] (f3),
      };
      \end{feynman} 
  \end{tikzpicture}
  \caption{}
  \label{fig:dipole_qg_t_channel}
  \end{subfigure}
  \begin{subfigure}{0.48 \textwidth}
    \centering
    \begin{tikzpicture}[scale=1.2,transform shape]
      \begin{feynman}
      \vertex (i1) {\(g\)};
      \vertex [crossed dot,below right = 1.41cm  of i1] (a){};
      \vertex [right =1.5cm of a] (b);
      \vertex [ right = 3cm of i2](c);
      \vertex [below right = 0.8cm of b](d);
      \vertex [below  =2cm of i1] (i2) {\(q_i\)};
      \vertex [ right =4cm of i1] (f1){\(q_j\)};
      \vertex [below = 1cm of f1] (f2){\(\nu\)};
      \vertex [below = 2cm of f1] (f3) {\( \bar \nu\)}; 
  
      \diagram* {
          (i1) -- [gluon](a),
          (i2) -- [fermion](a),
          (a) -- [fermion,edge label = \(q_j\)](b),
          (b) -- [ fermion] (f1),
          (d) -- [fermion] (f2),
          (d) -- [anti fermion] (f3),
          (b) -- [boson,edge label  = \(Z\)] (d),
          };
      \end{feynman}
      
  \end{tikzpicture}
  \caption{}
  \label{fig:dipole_qg_s_channel}
  \end{subfigure}
  \begin{subfigure}{0.48 \textwidth}
    \centering
   \begin{tikzpicture}[scale=1.2,transform shape]
        \begin{feynman}
          \vertex (i1) {\( q_i\)};
          
          \vertex [ crossed dot,right= 2cm of i1](a) {};
          \vertex [ below= 2cm of a](b);
          \vertex [right = 2cm of b] (c) ;
          \vertex [below  = 2cm of i1](i2) {\( \bar q_j\)};
          \vertex [right = 1cm of b](d);
          \vertex [ right= 2cm of a](f1){\(g\)};
          \vertex [ below = 1cm of f1](f3){\( \nu\)};
          \vertex [below  = 1cm of f3](f2){\(\bar\nu\)};
          \diagram* {
            (i1) -- [ fermion](a),
            (i2) -- [anti fermion](b),
            (a) -- [ fermion,edge label = \(q_j\)](b),
            (d) -- [anti fermion] (f2),
            (d) -- [fermion] (f3),
            (a) -- [gluon] (f1),
            (b) -- [boson,edge label = \(Z\)](d),
            };
        \end{feynman}
      \end{tikzpicture}
      \caption{}
\end{subfigure}
\begin{subfigure}{0.48 \textwidth}
  \centering
 \begin{tikzpicture}[scale=1.2,transform shape]
      \begin{feynman}
        \vertex (i1) {\( q_i\)};
        
        \vertex [ right= 2cm of i1](a) ;
        \vertex [crossed dot, below= 1.8cm of a](b) {};
        \vertex [right = 2cm of b] (c);
        \vertex [below  = 2cm of i1](i2) {\( \bar q_j\)};
        \vertex [right = 1cm of a](d);
        \vertex [ right= 2cm of b](f1){\(g\)};
        \vertex [ above = 1cm of f1](f3){\( \nu\)};
        \vertex [above  = 1cm of f3](f2){\(\bar\nu\)};
        \diagram* {
          (i1) -- [ fermion](a),
          (i2) -- [anti fermion](b),
          (a) -- [ fermion,edge label = \(q_i\)](b),
          (d) -- [anti fermion] (f2),
          (d) -- [fermion] (f3),
          (b) -- [gluon] (f1),
          (a) -- [boson,edge label = \(Z\)](d),
          };
      \end{feynman}
    \end{tikzpicture}
    \caption{}
    
\end{subfigure}

\caption{Parton level diagrams contributing to  $p p \rightarrow \nu \bar \nu + X$ with  insertions of gluonic dipole operators ($Q_{uG},Q_{dG}$)  at the  $q\bar qg$-vertex.
 The diagrams (a) and (b) show the contributions  from $q_ig \rightarrow q_j \nu_k \bar \nu_l$, while (c) and (d) the ones  from $q_i \bar q_j \rightarrow g \nu_k \bar \nu_l$.}
\label{fig:Dipole}
\end{figure}
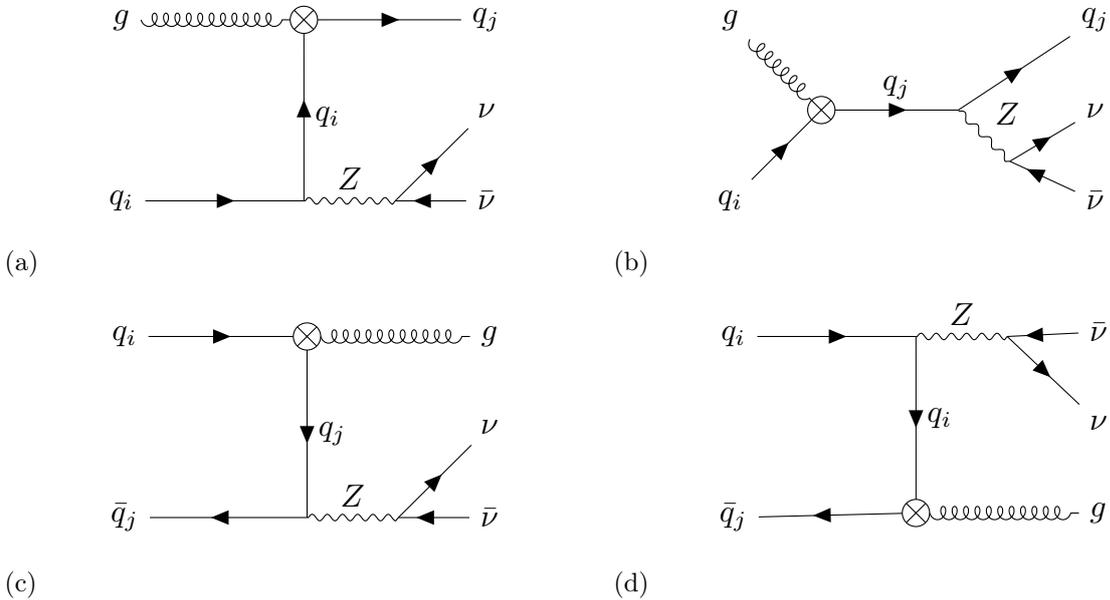
The chirality-flipped WC contributes to the same process, which corresponds to swapping flavor indices $i \Leftrightarrow j$.
The partonic BSM cross section  therefore depends on the effective WC 
\begin{equation}
  \label{eqn:gluon_Eff}
  C^{G}_{ij} = \sqrt{  |C_{qG,ij}|^2 + | C_{qG,ji}|^2 } ,
\end{equation}
where $q=u,d$ labels the type of quark, i.e., up-or down-type, which  is implicitly also  given by the flavor indices, say, up for $i=u,j=c$ and down for the other FCNCs.

\subsection{$Z$-vertex}
\label{sec:penguin}

Furthermore operator insertions modify the $q \bar q Z$ vertex and can therefore be constrained using the $pp \rightarrow \nu \bar \nu + X$ signature.
The effective Lagrangian is given by 
\begin{equation} \nonumber
  \begin{aligned}
  \mathcal{L}^{6}_{\text{SMEFT}} &\supset \frac{C^{(1)}_{\varphi q,ij}}{\Lambda_{NP}^2} ( \varphi^{\dagger} i \stackrel{\leftrightarrow}{D}_{\mu} \varphi ) (\bar q_i \gamma^{\mu} q_j ) +\frac{C^{(3)}_{\varphi q,ij}}{\Lambda_{NP}^2} ( \varphi^{\dagger} i \stackrel{\leftrightarrow}{D^I}_{\mu} \varphi ) \left(\bar q_i \tau^I \gamma^{\mu} q_j \right) \\
                      & + \frac{C_{\varphi u,ij} }{\Lambda_{NP}^2} ( \varphi^{\dagger} i \stackrel{\leftrightarrow}{D}_{\mu} \varphi ) (\bar u_i \gamma^{\mu}u_j) + \frac{C_{\varphi d,ij} }{\Lambda_{NP}^2} ( \varphi^{\dagger} i \stackrel{\leftrightarrow}{D}_{\mu} \varphi ) (\bar d_i \gamma^{\mu} d_j) \\
                      & +\frac{C_{uW,ij}}{\Lambda_{NP}^2} (\bar q_i \sigma^{\mu \nu} \tau^I  u_j  \tilde \varphi W^I_{\mu \nu}     ) + \frac{C_{dW,ij}}{\Lambda_{NP}^2}  (\bar q_i \sigma^{\mu \nu}  \tau^I  d_j \varphi W^I_{\mu \nu}     ) \\ 
                      & +\frac{C_{uB,ij}}{\Lambda_{NP}^2} (\bar q_i \sigma^{\mu \nu} u_j   \tilde \varphi B_{\mu \nu}     ) + \frac{C_{dB,ij}}{\Lambda_{NP}^2}  (\bar q_i \sigma^{\mu \nu}  d_j  \varphi B_{\mu \nu}     ) \, , 
                    \end{aligned}
\end{equation} 
and the corresponding diagrams with operator insertions are displayed in Fig.~\ref{fig:Zvertex}. 
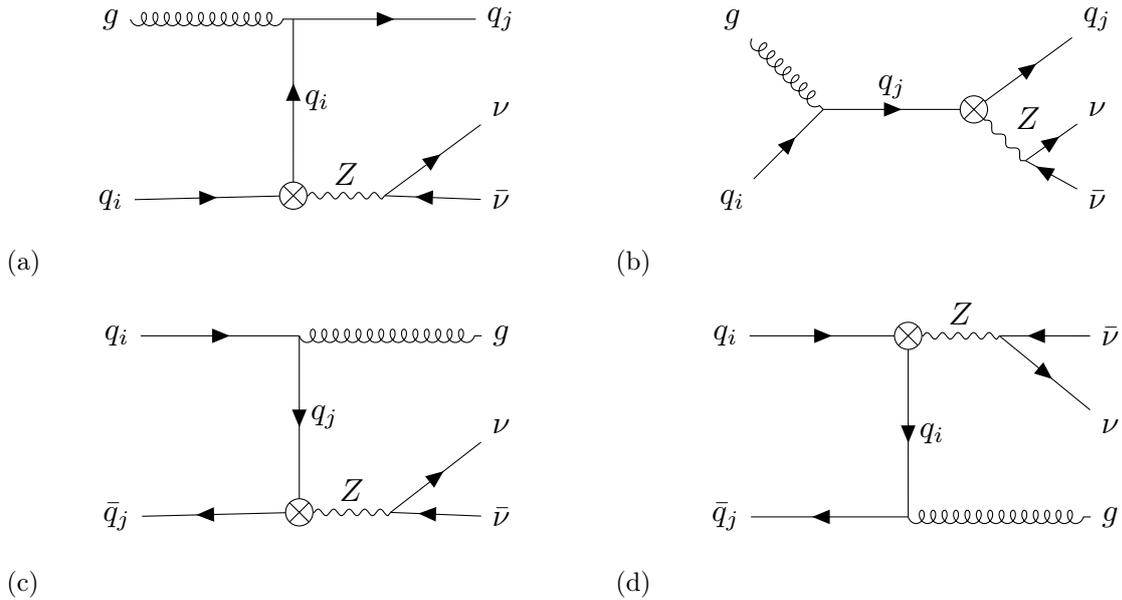
\begin{figure}[b]
  \centering
  \begin{subfigure}{0.48 \textwidth}
    \centering
    \begin{tikzpicture}[scale=1.2,transform shape]
      \begin{feynman}
      \vertex (i1) {\(g\)};
      \vertex [right = 2cm of i1] (a);
      \vertex [crossed dot,below = 1.8cm of a] (b){};
      \vertex [right = 2cm of b ] (c);
      \vertex [below = 2cm of i1] (i2) {\(q_i\)};
      \vertex [right = 1cm of b] (d);
      \vertex [right = 2cm of a] (f1){\(q_j\)};
      \vertex [below = 1cm of f1] (f2){\(\nu\)};
      \vertex [below = 2cm of f1 ] (f3) {\( \bar \nu\)}; 
      \diagram* {
      (i1) -- [gluon](a),
      (i2) -- [fermion](b),
      (a) -- [anti fermion,edge label = \(q_i\)](b),
      (a) -- [ fermion] (f1),
      (b) -- [ boson,edge label = \(Z\)] (d),
      (d) -- [fermion] (f2),
      (d) -- [anti fermion] (f3),
      };
      \end{feynman} 
  \end{tikzpicture}
  \caption{}
  \label{fig:dipole_qg_t_channel}
  \end{subfigure}
  \begin{subfigure}{0.48 \textwidth}
    \centering
    \begin{tikzpicture}[scale=1.2,transform shape]
      \begin{feynman}
      \vertex (i1) {\(g\)};
      \vertex [below right = 1.41cm  of i1] (a);
      \vertex [crossed dot,right =1.5cm of a] (b){};
      \vertex [ right = 3cm of i2](c);
      \vertex [below right = 0.8cm of b](d);
      \vertex [below  =2cm of i1] (i2) {\(q_i\)};
      \vertex [ right =4cm of i1] (f1){\(q_j\)};
      \vertex [below = 1cm of f1] (f2){\(\nu\)};
      \vertex [below = 2cm of f1] (f3) {\( \bar \nu\)}; 
  
      \diagram* {
          (i1) -- [gluon](a),
          (i2) -- [fermion](a),
          (a) -- [fermion,edge label = \(q_j\)](b),
          (b) -- [ fermion] (f1),
          (d) -- [fermion] (f2),
          (d) -- [anti fermion] (f3),
          (b) -- [boson,edge label  = \(Z\)] (d),
          };
      \end{feynman}
      
  \end{tikzpicture}
  \caption{}
  \label{fig:dipole_qg_s_channel}
  \end{subfigure}
  \begin{subfigure}{0.48 \textwidth}
    \centering
   \begin{tikzpicture}[scale=1.2,transform shape]
        \begin{feynman}
          \vertex (i1) {\( q_i\)};
          
          \vertex [ right= 2cm of i1](a) ;
          \vertex [crossed dot, below= 1.8cm of a](b){};
          \vertex [right = 2cm of b] (c) ;
          \vertex [below  = 2cm of i1](i2) {\( \bar q_j\)};
          \vertex [right = 1cm of b](d);
          \vertex [ right= 2cm of a](f1){\(g\)};
          \vertex [ below = 1cm of f1](f3){\( \nu\)};
          \vertex [below  = 1cm of f3](f2){\(\bar\nu\)};
          \diagram* {
            (i1) -- [ fermion](a),
            (i2) -- [anti fermion](b),
            (a) -- [ fermion,edge label = \(q_j\)](b),
            (d) -- [anti fermion] (f2),
            (d) -- [fermion] (f3),
            (a) -- [gluon] (f1),
            (b) -- [boson,edge label = \(Z\)](d),
            };
        \end{feynman}
      \end{tikzpicture}
      \caption{}
\end{subfigure}
\begin{subfigure}{0.48 \textwidth}
  \centering
 \begin{tikzpicture}[scale=1.2,transform shape]
      \begin{feynman}
        \vertex (i1) {\( q_i\)};
        
        \vertex [ crossed dot,right= 2cm of i1](a) {};
        \vertex [ below= 2cm of a](b) ;
        \vertex [right = 2cm of b] (c);
        \vertex [below  = 2cm of i1](i2) {\( \bar q_j\)};
        \vertex [right = 1cm of a](d);
        \vertex [ right= 2cm of b](f1){\(g\)};
        \vertex [ above = 1cm of f1](f3){\( \nu\)};
        \vertex [above  = 1cm of f3](f2){\(\bar\nu\)};
        \diagram* {
          (i1) -- [ fermion](a),
          (i2) -- [anti fermion](b),
          (a) -- [ fermion,edge label = \(q_i\)](b),
          (d) -- [anti fermion] (f2),
          (d) -- [fermion] (f3),
          (b) -- [gluon] (f1),
          (a) -- [boson,edge label = \(Z\)](d),
          };
      \end{feynman}
    \end{tikzpicture}
    \caption{}
    
\end{subfigure}
\caption{Parton level diagrams contributing to  $p p \rightarrow \nu \bar \nu + X$ with  insertions of operators at the $q\bar q Z$-vertex. It  can be modified by penguin operators ($ Q^{(1)}_{\varphi q},Q^{(3)}_{\varphi q},Q_{\varphi u} ,Q_{\varphi d}$) and EW dipole operators ($Q_{uW},Q_{dW},Q_{uB},Q_{dB}$). 
 The diagrams (a) and (b) show the contributions  from $q_ig \rightarrow q_j \nu_k \bar \nu_l$, while (c) and (d) the ones  from $q_i \bar q_j \rightarrow g \nu_k \bar \nu_l$.
 }
\label{fig:Zvertex}
\end{figure}

The operators belong to two categories: The  penguin operators ($Q^{(1)}_{\varphi q},Q^{(3)}_{\varphi q}, Q_{\varphi u}, Q_{\varphi u}$) with vector currents and the EW dipole operators ($Q_{uB},Q_{dB},Q_{uW},Q_{dW}$) with tensorial structure.

For the EW dipole operators the effective WC is given by 

\begin{equation}
  \label{eqn:EW_dip}
 {C^{EW}_{ij}} = \sqrt{ |\cos \theta_W C_{qW,ij}  \mp \sin \theta_W C_{qB,ij}|^2 +|\cos \theta_W C_{qW,ji} \mp \sin \theta_W C_{qB,ji}|^2 },
\end{equation}
where $q=u,d$ labels the type of quarks, with the relative $- (+)$  sign corresponding to up (down)-type ones, and $\theta_W$ is the weak mixing angle.
As  already argued in Sec.~\ref{sec:gluon-dipole}  for the gluon dipole operators also  for the electroweak dipoles  the chirality flipped WC also contributes
and is included in \eqref{eqn:EW_dip}.

For the penguin operators the situation is more involved due  to the $SU(2)$-link  of the  doublet quark currents, which
mixes up- and down-sector quarks.
Denoting here quark fields  in components $q=(u_L,d_L)$  in the mass basis by a prime, $u_L =V_u u_L^\prime$, $d_L =V_d d_L^\prime$, with unitary  matrices $V_{u,d}$,
 the CKM-matrix is given as $V=V_u^\dagger V_d$, and 
\begin{align}
    \begin{aligned}  \label{eq:CO}
    C_{\varphi q, ij }^{(1)} Q_{\varphi q,ij}^{(1)} &= C_{\varphi q,ij}^{(1)} \left(\varphi^\dagger i\overleftrightarrow D_\mu  \varphi\right)\left(\bar u_L^i \gamma^\mu u_L^j  + \bar d_L^i \gamma^\mu d_L^j\right)\\ 
    &=C_{\varphi q,ij}^{(1)} \left(\varphi^\dagger i\overleftrightarrow D_\mu  \varphi\right)\left(\left(V^{\dagger}_u\right)_{ki}\bar u_L^{\prime k} \gamma^\mu \left(V^{\vphantom{\dagger}}_u\right)_{jl} u_L^{\prime l}  + \left(V^{\dagger}_d\right)_{mi}\bar d_L^{\prime m} \gamma^\mu \left(V^{\vphantom{\dagger}}_d\right)_{jn} d_L^{\prime n}\right)\\
    &=    C_{\varphi q,ij}^{(1)} \left(V^{\dagger}_u\right)_{ki} \left( V^{\vphantom{\dagger}}_u\right)_{jl}  \left(\varphi^\dagger i\overleftrightarrow D_\mu  \varphi\right)\left(\bar u_L^{\prime k} \gamma^\mu  u_L^{\prime l} + V^\dagger_{mk} V^{\vphantom{\dagger}}_{ln} \bar d_L^{\prime m} \gamma^\mu  d_L^{\prime n}\right) \, , 
    \end{aligned}
\end{align}
with similar  expression  for $Q_{\varphi q,ij}^{(3)}$ with opposite relative sign between the currents.
Here, we  collect all CKM-factors in front of the down-type currents, suitable for the up-mass basis, in which we spell out limits on up-type FCNCs.
We can do the analogous  thing in  the down-mass basis, to work out limits on down-type FCNCs.
From (\ref{eq:CO}) follows that switching on a
$|\Delta c|=|\Delta u|=1$ FCNC contribution  ($k=1,l=2$) leads simultaneously to down-type operators,
schematically,
\begin{align} \label{eq:extra}
\bar u_L^\prime \gamma_\mu  c_L^\prime ,  \quad \bar d_L^\prime  \gamma_\mu s_L ^\prime ,  \quad O(\lambda)-\mbox{terms} \, , 
\end{align}
in addition to terms of higher order in the Wolfenstein parameter $\lambda \simeq 0.2$, which reflects the  hierarchies in the CKM matrix.
For FCNCs with beauty-quarks, operators with tops are present, 
\begin{align}
\bar s_L ^\prime   \gamma_\mu b_L^\prime,  \quad  \bar c_L^\prime   \gamma_\mu  t_L^\prime, \quad O(\lambda^2)-\mbox{terms} , \\
\quad \bar d_L^\prime   \gamma_\mu  b_L^\prime,  \quad  \bar  u_L^\prime    \gamma_\mu t_L ^\prime, \quad O(\lambda^3)-\mbox{terms}  \, .
\end{align}
However, since tops do not contribute to Drell-Yan production, one can provide bounds on single flavor  combinations.

Taking this into account allows to constrain,
with flavors made explicit, 
\begin{eqnarray}
  \label{eqn:WC_eff_penguin-SU2}
   C^{ZP}_{uc} & =\sqrt{ |C^{(-)}_{\varphi q,uc}|^2 + |C_{\varphi u,uc}|^2 + r_{ds}  |C^{(+)}_{\varphi q,ds}|^2 }, \\
    C^{ZP}_{ds} & =\sqrt{ |C^{(+)}_{\varphi q,ds}|^2 + |C_{\varphi d,ds}|^2 + r_{uc}  |C^{(-)}_{\varphi q,uc}|^2 }, \nonumber
\end{eqnarray}
where $r_{ds}$ ($r_{uc}$)  is the relative weight after PLF-folding to the cross section from the induced $ds$ ($uc$) operator. Note,
$r_{ds} <  r_{uc}=1/r_{ds}$, and both are not far from  one due to the proximity of the PLFs. (For the recast parameters, cf Sec.~\ref{sec:simulation},  one obtains 
 the net values $r_{ds}\simeq 0.69$ and $r_{uc} \simeq 1.45$, but note that the coefficients depend on the kinematics and bins). 
Note also that we neglected  in (\ref{eq:extra}) the diagonal contributions from quarks of the first two generations because on the level of cross section their impact is
suppressed by $\lambda^2$, while none of them is enhanced at that level compared to the FCNC ones (see  Fig.~\ref{fig:PLFs}).
For $i=d,s$ and $j=b$ the coefficient
\begin{equation}
  \label{eqn:WC_eff_penguin}
   C^{ZP}_{ij} =\sqrt{ |C^{(\mp)}_{\varphi q,ij}|^2 + |C_{\varphi u/d,ij}|^2 },
\end{equation}
 is sufficient.
Here and in (\ref{eqn:WC_eff_penguin-SU2}) 
\begin{equation}
  C^{(\mp)}_{\varphi q,ij} = C^{(1)}_{\varphi q,ij} \mp C^{(3)}_{\varphi q,ij} \, , 
\end{equation}
where the  '$-$' and $u$ is for up-type quarks and the '$+$' and $d$ for down-type quarks~\footnote{ \label{foot:C13} To understand the  sign differences between the singlet and the triplet coefficient, recall that the covariant derivative  for the $Z$-couplings is proportional to $\sigma^{I=3} W^{I=3}_\mu$.}.

\section{Missing Energy Observables at the LHC \label{sec:Emiss}}

We discuss the missing transverse energy distribution ($E_T^{miss}$)  in association with an  energetic jet  in 
the process $ pp \rightarrow X + \text{invisible}$, where $X$ denotes hadronic final states, that result from the jet after showering  and hadronization.
The missing transverse momentum vector is defined as 
\begin{equation}
  \label{eqn:ETmiss}
  \bold{p}_T^{miss} = -\sum_i \bold{p}_T^i,  \quad \quad E_T^{miss}=| \bold{p}_T^{miss} | \, , 
\end{equation}
where the sum includes the transverse momentum vectors of all visible final particles, and 
the scalar quantity $E_T^{miss}$ is defined as the magnitude of $\bold{p}_T^{miss}$.
The  jet corresponds here to the leading jet   with large modulus of  transverse momentum, $P_T \gtrsim \SI{150}{\GeV}$.
The leading jet is usually accompanied by several subleading jets ($P_T \gtrsim \SI{30}{\GeV}$),  all of which contribute to Eq.~\eqref{eqn:ETmiss}.

We work out  the parton luminosities for quark-antiquark and quark-gluon fusion  in Sec.~\ref{sec:lumis} and
discuss the energy enhancement of the SMEFT operator contributions to the  missing energy spectra  in Sec.~\ref{sec:shapes}.
Details on the analytical calculation are  given in  App.~\ref{app:pertub}. 
The recast of the ATLAS search  \cite{ATLAS:2021kxv}  is given in Sec.~\ref{sec:simulation}.

\subsection{Proton structure and missing energy spectrum \label{sec:lumis}}

Generally $\nu \bar \nu$ production is described by the Drell-Yan process \cite{Drell:1970wh}, however the process $q \bar q \rightarrow \nu \bar \nu$ does not generate any $P_T$ and therefore the leading order (LO) contributions involve an  additional  energetic  quark or gluon.
Explicitly these are given by $ q \bar q \rightarrow \nu \bar \nu g $, $ q g \rightarrow \nu \bar \nu q $ and $ \bar q g \rightarrow \nu \bar \nu \bar q $, where the former two are related by crossing symmetry and the latter two are related through charge conjugation.
Note that the flavor of final state neutrinos are incoherently summed over, since the experimental analysis is blind to neutrino flavors.
The hadronic cross section can be written as 
\begin{equation}
  \label{eqn:E_miss}
  \frac{\mathrm{d}\sigma }{\mathrm{d}E_T^{miss}} = \sum_{i,j}\int \frac{\mathrm{d}\tau }{\tau}   \left\{ \frac{\mathrm{d} \hat \sigma_{q_i \bar q_j}(\tau s,E_T^{miss}) }{\mathrm{d}E_T^{miss}  } \mathcal{L}_{i  j}(\tau) + \frac{\mathrm{d} \hat \sigma_{q_i g}(\tau s,E_T^{miss}) }{\mathrm{d}E_T^{miss} } \mathcal{L}_{i g}(\tau)     \right\},
\end{equation}
where $\tau = \frac{\hat s}{s}$ is the ratio of partonic  and hadronic  center-of-mass energy-squared, the sum runs over flavors $u,c,d,s,b$  and  at this order of perturbation theory holds $E_T^{miss} = P_T$, where $P_T$ is the transverse momentum of the  final state particle $X$. 
Hard cross sections $\mathrm{d} \hat \sigma_{q_i \bar q_j}$ and $\mathrm{d} \hat \sigma_{q_i g}$ for the processes $ q_i \bar q_j \rightarrow \nu \bar \nu g $ and $ q_i g \rightarrow \nu \bar \nu q_j $ can be calculated in perturbation theory.
Note that the process $ \bar q g \rightarrow \nu \bar \nu \bar q $ is  included through the summation over all partons and the $ \mathcal{L}_{i  j}$ functions.
The proton structure in Eq.~\eqref{eqn:E_miss} is captured through the parton luminosity functions (PLFs), which are defined as 
\begin{equation}
    \label{eqn:PLFs}
    \mathcal{L}_{i j}(\tau) = \tau \int_{\tau}^{1} \frac{\mathrm{d}x}{x} \left[ f_i(x,\mu_F) f_{\bar j}(\tau/x,\mu_F) + f_j(x,\mu_F) f_{\bar i}(\tau/x,\mu_F)  \right],
  \end{equation} 
where $i,j = u,d,s,c,b,g$ with $\bar g = g$ and $f_i(x,\mu_F)$ denote the PDFs with the factorization scale $\mu_F$.
The PLFs essentially capture the rate of initiation of processes between two partons and are displayed in Fig.~\ref{fig:PLFs}.
\footnote{We use here   $\mu_F = \sqrt{\hat s}$ to allow for better display of  hierarchies, however the recast  is done with
$\mu_F = \sum_i P_{T,i}/2$. \label{foot:scale}}
\begin{figure}
    \begin{subfigure}{0.48\textwidth}
      \includegraphics[width = \textwidth]{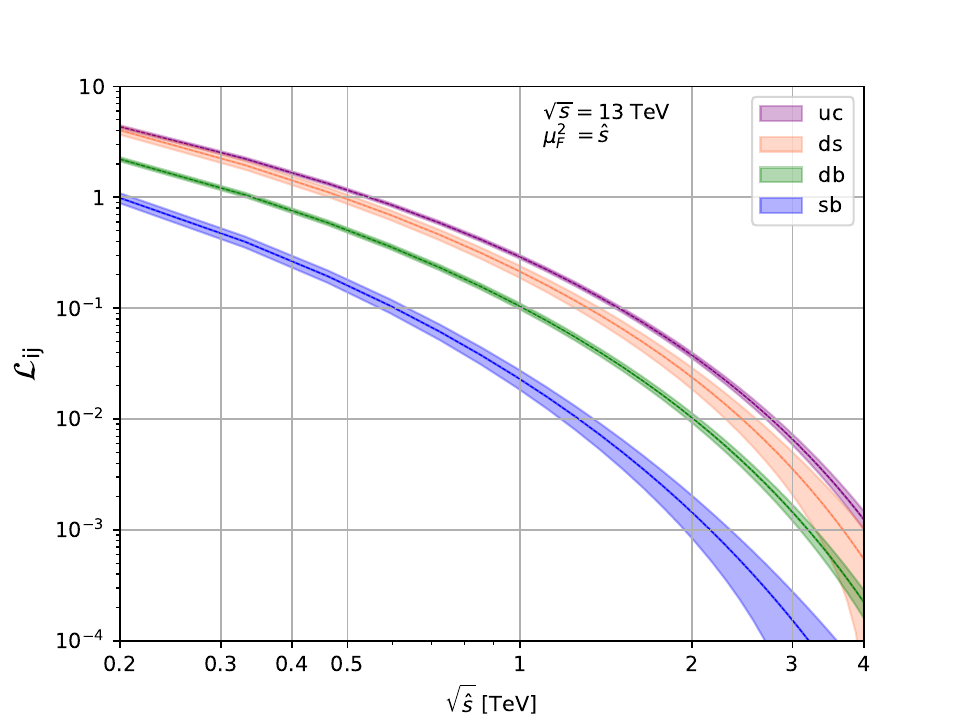}
    \end{subfigure}
    \begin{subfigure}{0.48\textwidth}
      \includegraphics[width = \textwidth]{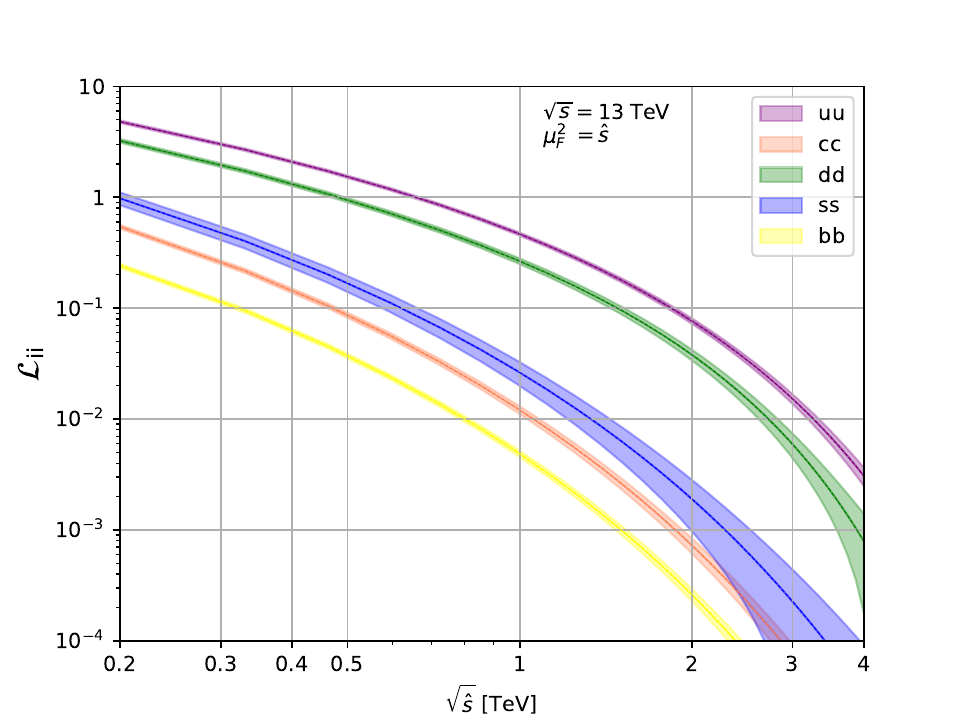}
    \end{subfigure}
    \begin{subfigure}{0.48\textwidth}
      \includegraphics[width = \textwidth]{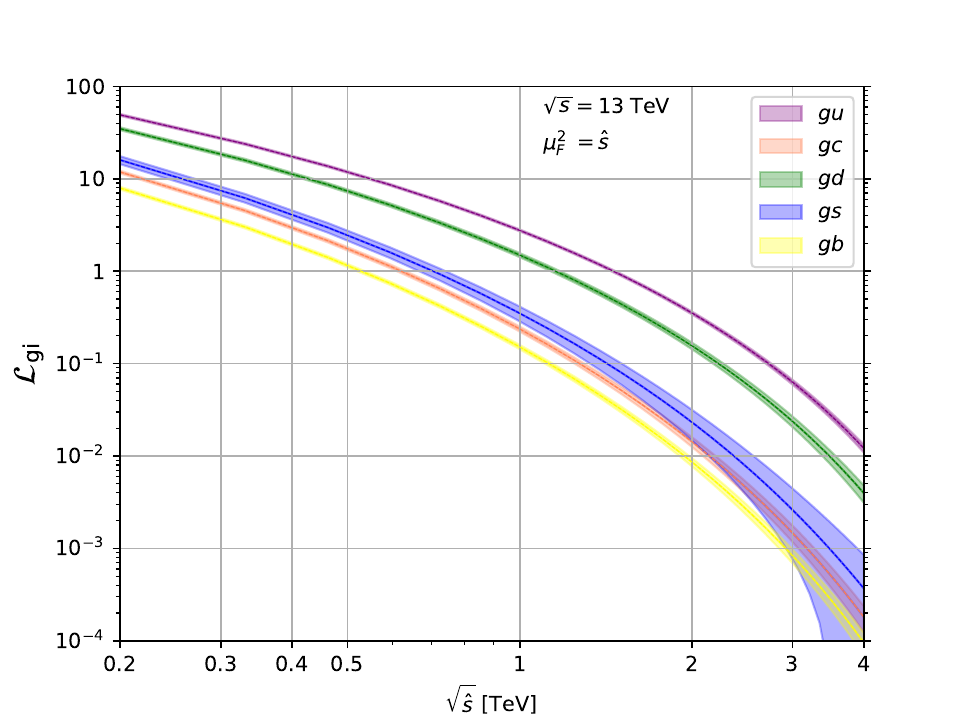}
    \end{subfigure}
    \caption{The parton luminosity functions \eqref{eqn:PLFs} for different combinations of partons. The upper left panel  corresponds the  flavor off-diagonal quark combinations and the upper right figure to  flavor diagonal quark combinations. The lower plot shows the combinations of quarks and gluons. We use the PDFset $\texttt{PDF4LHC15\_nlo\_mc}$ \cite{PDFset,PDFset2,PDFset3,PDFset4} with the  factorization scale  $\mu_F = \sqrt{\hat s}$. Solid lines correspond to the central values and the envelope shows the $1 \sigma$-ranges for the PDF uncertainties.}
    \label{fig:PLFs}
  \end{figure}
Note that the definition of the PLFs  (\ref{eqn:PLFs}) differs  from the one in Ref.~\cite{Angelescu:2020uug}, since we group together $q\bar q'$ and $q'\bar q$, since this definition is manifestly symmetric under $i \Leftrightarrow j$ and we extend the definition to include gluons.
On the other hand  for diagonal quark combinations equation \eqref{eqn:PLFs} is identical with the one in  Ref.~\cite{Angelescu:2020uug} and, moreover, only slight differences  for
 off-diagonal couplings are visible.
Furthermore the total cross section is  identical, regardless of the definition, since the sum over all possible inital partons has to be considered.
  
  The partonic cross sections are  worked out analytically, to understand the shape and relative size of the NP contributions, as well as to validate the
simulation based on tools decribed in Sec.~\ref{sec:simulation}.
This framework generalizes  the parametrization used in \cite{Lam:1978pu} and can  be extended to other processes, such as $e^+ e^- \rightarrow \text{Hadrons}, e^+ p \rightarrow X, p p \rightarrow \ell^{-} \ell^{+} + X$.
Details  are provided  in App.~\ref{app:pertub}.

  \subsection{Shapes of SMEFT-distributions  \label{sec:shapes}}
  
\begin{figure}
  \centering
  \includegraphics[width = \linewidth]{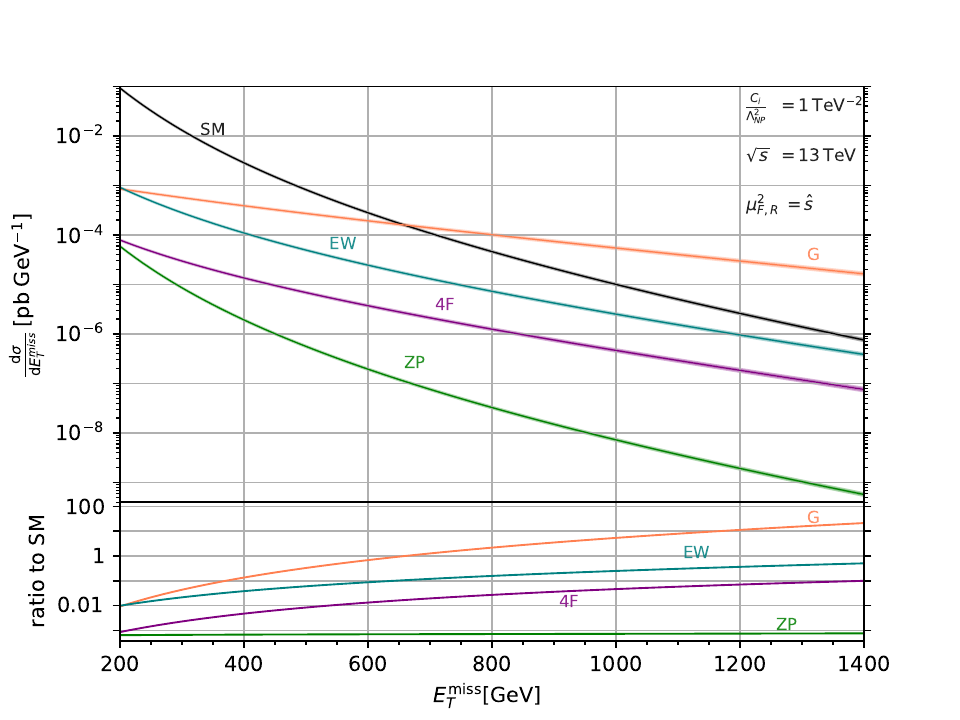}
  \caption{ $E_T^{miss}$ distributions at LO for $pp \rightarrow \nu \bar \nu + X$ in the SM (black) and  insertions of FCNC
  gluon dipole ($G$, orange), electroweak dipole ($EW$, petrol), four-fermion ($4F$, pink) and $Z$-penguin ($ZP$,  green) operators
  inducing $uc$-transitions. NP effects are for $\Lambda_{NP} = \SI{1}{\TeV}$ and effective WCs fixed to $C^i_{uc} =1$ and all others to zero,
  to illustrate the relative shapes and reach of the  NP contributions, see text.  We use the PDFset $\texttt{PDF4LHC15\_nlo\_mc}$ \cite{PDFset,PDFset2,PDFset3,PDFset4}
  and $\mu_F, \mu_R=\sqrt{\hat s}$. PDF uncertainties are included, yet very small. }
   \label{fig:xsec}
\end{figure}

We show in  Fig.~\ref{fig:xsec} the  LO  $E_T^{miss}$ differential cross sections  after PLF-folding for the SM and exemplarily  for operator insertions with $uc$ FCNCs,
see also footnote \ref{foot:scale}.
Fig.~\ref{fig:xsec} illustrates  the shapes and hierarchies between  different operator insertions. The  effect from flavor can be inferred from the hierarchies of the PLFs, shown 
 in Fig.~\ref{fig:PLFs}.
  The lower part of  Fig.~\ref{fig:xsec}  gives the ratio of NP spectra to the SM one, displaying the relative energy growth of the different operators.
 Since $\Lambda_{NP}/\sqrt{C_{uc}^i}= \SI{1}{\TeV}$ is a rather low value,  this should be considered  as an illustration rather than the typical NP benchmark.

Since the $Z$-penguins  ($ZP$, green) have the same Dirac-structure as the SM (black), the shapes of their $P_T$-distributions are identical. The other contributions experience further energy-enhancements.
Recall that both quark-antiquark annihilation as well as quark-gluon fusion diagrams contribute, which are folded with different PLFs.
At parton level, quark-antiquark fusion has always the larger contributions except for the gluon dipoles, for which they are of similar size.
On the other hand, the $qg$-PLFs are about one order of magnitude larger. After folding,
the largest impact has the gluon dipole ($G$, orange), followed by the EW dipole ($EW$, petrol) and four-fermion operators ($4F$, pink).

The shapes and hierarchies  in  Fig.~\ref{fig:xsec}  can be understood from the differential partonic cross sections,
here for $ q_i g \rightarrow \nu \bar \nu q_j$. In the high energy limit, $M_Z,v \ll P_T \sim \sqrt{\hat s}/2$, they can be written as
\begin{align} \label{eq:parton-SM}
\frac{\mathrm{d}\hat \sigma_{SM}( q_i g \rightarrow \nu \bar \nu q_j)}{\mathrm{d} P_T }      &\approx  \frac{ \alpha_s  \text{Br}( Z \rightarrow \nu \bar \nu) \left( {\epsilon_L^{i,j}}^2 + {\epsilon_R^{i,j}}^2   \right)}{12 \sqrt{2}}\frac{1}{\hat s^{3/2} } \frac{ x^2 + 4 }{ x \sqrt{ 1- x^2} }  \, , 
 \\ \label{eq:parton-ZP}
\frac{\mathrm{d} \hat \sigma_{ZP}( q_i g \rightarrow \nu \bar \nu q_j)}{\mathrm{d} P_T } &\approx \frac{ \alpha_s{C^{ZP}_{ij}}^2 \text{Br}\left( Z \rightarrow \nu \bar \nu\right)}{24 }  \frac{ M_Z^2 v^2}{ \Lambda_{NP}^4 } \frac{1}{\hat s^{3/2}} \frac{  x^2+4}{ x \sqrt{1-x^2}} \, , 
\\ \label{eq:parton-4F}
\frac{\mathrm{d}\hat \sigma_{4F}( q_i g \rightarrow \nu \bar \nu q_j)}{\mathrm{d} P_T }      &\approx   \frac{5\alpha_s {C^{4F}_{ij} }^2}{432 \sqrt{2} \pi^2 } \frac{1}{\Lambda_{NP}^4}\sqrt{\hat s} (1-x)^{3/2}  \, ,   \\   \label{eq:parton-EW}
\frac{\mathrm{d} \hat \sigma_{EW}( q_i g \rightarrow \nu \bar \nu q_j)}{\mathrm{d} P_T }     &\approx  \frac{ \alpha_s (C^{EW}_{ij})^2 \text{Br}\left( Z \rightarrow \nu \bar \nu\right) }{6 \sqrt{2}  }  \frac{v^2}{\Lambda_{NP}^4} \frac{1}{\sqrt{\hat s} } \frac{ x}{\sqrt{ 1- x^2}} \, ,  \\ 
\frac{\mathrm{d} \hat \sigma_{G}( q_i g \rightarrow \nu \bar \nu q_j)}{\mathrm{d} P_T } &\approx \frac{ {C^{G}_{ij}}^2  \text{Br}\left( Z \rightarrow \nu \bar \nu\right)\left(\epsilon_L^{i,j}- {\epsilon_R^{i,j} }\right)^2 v^2}{96 \pi M_Z^2  } \frac{1}{\Lambda_{NP}^4} \sqrt{\hat s} \frac{x  }{\sqrt{ 1- x^2} } \, ,   \nonumber \\
&= \frac{ {C^{G}_{ij}}^2  \text{Br}\left( Z \rightarrow \nu \bar \nu\right) }{96 \pi  } \frac{1}{\Lambda_{NP}^4} \sqrt{\hat s} \frac{x  }{\sqrt{ 1- x^2} } \, ,   \label{eq:parton-G}
\end{align}
where $x = 2 P_T \, / \sqrt{\hat s}$, $\text{Br}( Z \rightarrow \nu \bar \nu)\simeq 20 \%$ \cite{ALEPH:2005ab} is the SM branching ratio of the $Z$-boson to neutrinos, and  $\epsilon_{L(R)}^{i,j}$  are the SM $Z$-couplings to left-(right-) handed quarks \eqref{eqn:SM_Z_couplings}. 
 Here, we employed the narrow-width approximation, but used the full expressions in the numerical analysis.
The energy-dependence of the quark-antiquark fusion processes is similar, and given in App.~\ref{app:pertub}.
Due to the larger PLFs  the hadronic $P_T$-spectra are dominated by quark-gluon fusion. 

Comparing the differential cross section of the $Z$-penguins (\ref{eq:parton-ZP}) with the SM one, (\ref{eq:parton-SM}), which drops as $1/\hat s^{3/2}$, one observes  that the former is suppressed by 
$\sim v^4/\Lambda_{NP}^4$. On the other hand, the EW-dipoles (\ref{eq:parton-EW})  are lesser suppressed by the NP scale, and receive partial energy enhancement
relative to the SM as $\sim v^2 \hat s /\Lambda_{NP}^4$. Both four-fermion  (\ref{eq:parton-4F}) and gluon dipole operators  (\ref{eq:parton-G}) are fully energy-enhanced relative to the SM, $\sim \hat s^2/\Lambda_{NP}^4$. 
All but the four-fermion insertion involve the $Z$-resonance to produce dineutrinos, therefore only the 4F-process is  a  genuine $(2 \rightarrow 3)$-process, and has stronger
phase space suppression than the others including the SM.

\section{Recast of the experimental analysis}
\label{sec:simulation}

In this section the simulation chain used for the recast of the experimental analysis  \cite{ATLAS:2021kxv}  is described.
First the feynman rules for the process are exported as an UFO model, which is then read in by \texttt{MadGraph5\_aMC@NLO} \cite{MG5}.
For the SM calculations the "sm" UFO file is used, while the SMEFT contributions are based on the UFO model \texttt{SMEFTsim\_general\_MwScheme\_UFO} \cite{SMEFTsim}.
Using \texttt{MadGraph5\_aMC@NLO} statistical significant events samples of $ pp \rightarrow \nu \bar \nu + X$ are generated. 
The resulting samples are showered and hadronized using \texttt{Pythia8} \cite{pythia} and finally detector effects are estimated using \texttt{Delphes3} \cite{delphes}.
For this the ATLAS Delphes card is used and final selection criteria are applied using \texttt{ROOT}. Jets are clustered with the anti-$k_T$ algorithm with a jet radius $\Delta R = 0.4$ using the program \texttt{fastJet} \cite{fastjet}.
The factorization and renormalization scales are chosen as $\mu_R = \mu_F = \frac{1}{2}\sum_i P_{T,i}$, where the sum is over all final states. 
  We use the PDFset $\texttt{PDF4LHC15\_nlo\_mc}$ \cite{PDFset,PDFset2,PDFset3,PDFset4}.

We recast the ATLAS analysis \cite{ATLAS:2021kxv}, which examines events with an energetic jet and missing transverse momentum. The dataset is based on Run II data and corresponds to an integrated luminosity $\mathcal{L}_{int} = \SI{139}{\femto \barn}^{-1}$ at a center-of-mass energy $\sqrt{s} = \SI{13}{\TeV}$.
First SM background samples are  calculated and compared to the background analysis reported in \cite{ATLAS:2021kxv}. The results are reproduced to an accuracy of $\SI{10}{\percent}-\SI{20}{\percent}$.
Subsequently the pure BSM contribution  is computed for  $ C_i = 1$ and then rescaled to obtain limits.

We also work out  projections  for the HL-LHC ($\mathcal{L}_{int} = \SI{3000}{\femto\barn}^{-1} $), assuming that data scales with the luminosity ratio, while all uncertainties scale with the square root of the luminosity ratio.
These projections can be considered as rather conservative, since the HL-LHC will increase the center-of-mass energy to $\sqrt{s} = \SI{14}{\TeV}$. This  could lead to additional bins for high-$P_T$  in the tails of the distribution, that are currently not probed, however are  expected to be most sensitive.

The constraints on the effective WCs are extracted using the CLs method \cite{CLs}. These are computed using the framework \texttt{pyhf} \cite{pyhf} and all WCs with $\text{CLs} < 0.05$ are excluded.  
We estimated theory  uncertainties from PDF and scale variation, and although they are sizeable, around 20 \%  for the highest bin,
in absolute terms this is negligible compared to the background uncertainties.
For the $uc$ quark flavor we also considered PDFs with intrinsic charm \cite{NNPDF:2023tyk}, which result in  about 10 \% larger cross sections.
 Therefore, systematic errors for the signal simulations are neglected. 
We validate our numerical simulation with a study of analytical properties, which is detailed in App.~\ref{app:pertub}.
Missing energy spectra based on Madgraph and our analytical computation are  compared in Fig.~\ref{fig:validation}, and show very good agreement with each other.
The input parameters for the analytical results are based on Ref.~\cite{ParticleDataGroup:2022pth}.
 We considered all available bins in the analysis, however, we verified that  the fit is dominated by the highest three bins.
The results of the recast are presented in Sec.~\ref{sec:bounds}.

\section{Constraining New Physics  \label{sec:bounds}}

We give  the results of the SMEFT  recast of the $pp \to \text{MET}+j$ search \cite{ATLAS:2021kxv}.
The limits, presented in Sec.~\ref{sec:results}, are obtained  by switching on a single effective NP coupling, while setting all others to zero.
For operators containing left-handed  up-type (down-type) quarks we present limits in the up-mass (down-mass) basis.
Our analysis also allows for an application to models with light right-handed neutrinos, discussed  in Sec.\ref{sec:RHN}.
In Sec~\ref{sec:comp} we compare our  constraints on  semileptonic four-fermion operators with those from rare decays and Drell-Yan  production.
We make  comments on  dipole coefficients and compare available constraints  in Sec.~\ref{sec;dipole-comp}.

\subsection{Results for  SMEFT operators \label{sec:results}}

\begin{table}[h]
    \centering
    \begin{tabular}{c|c|c|c|c}
    \toprule
     ${ij}$         & $C^{G}_{ij}$   & $C^{EW}_{ij}$     & $C^{4F}_{ij}$  & $C^{ZP}_{ij}$ \\
    \midrule
     $uc$       &  \num{0.0077}(\num{0.0059})    &  \num{0.036} (\num{0.029})    &  \num{0.079}(\num{0.060})  &   \num{0.39} (\num{0.30})    \\
     $ds$       &    \num{0.011} (\num{0.009}) &  \num{0.051} (\num{0.040})   &  \num{0.11} (\num{0.09})   &   \num{0.47} (\num{0.34})  \\
     $db$       &   \num{0.012} (\num{0.010})   &  \num{0.056} (\num{0.043})  &  \num{0.14} (\num{0.10})    &   \num{0.54} (\num{0.38})  \\
     $sb$        &  \num{0.026} (\num{0.020})  &  \num{0.12} (\num{0.10}) &  \num{0.37} (\num{0.30})   &    \num{1.07} (\num{0.78})  \\
    \bottomrule
    \end{tabular} 

    \caption{ The $95\%$ CL-limits on  effective  gluon dipole WCs \eqref{eqn:gluon_Eff},   electroweak  dipole WCs \eqref{eqn:EW_dip}, four-fermion WCs \eqref{eqn:WC_eff},  and $Z$-penguins \eqref{eqn:WC_eff_penguin} for off-diagonal quark flavor combinations $i \neq j$ and a NP scale $\Lambda_{NP} = \SI{1}{\TeV}$ from the ATLAS search  \cite{ATLAS:2021kxv}. Values in parentheses correspond to  projections ($ \mathcal{L}_{int} = \SI{3000}{\femto\barn}^{-1}$) assuming naive statistical scaling.  } 
    \label{tab:fcnc_bounds} 
  \end{table}

In  Table \ref{tab:fcnc_bounds} we give the bounds  on the effective coefficients and projections  for the high-luminosity upgrade 
for the FCNC semileptonic four-fermion, gluonic dipole,  EW dipole operators and $Z$-penguin operators.
As the parton luminosities are the largest for $uc$, followed by $ds$, $db$ and $sb$, see Fig.~\ref{fig:PLFs}, the bounds are strongest for
$uc$, followed by $ds$, $db$ and $sb$. 
 Note,  intrinisic charm PDFs  would  mildly increase  $uc$ cross sections, and  therefore  the corresponding limits  in  Table \ref{tab:fcnc_bounds} 
would strengthen by about 5 \%.
Limits can be translated into a NP scale $\Lambda_{NP}$  that can be probed,  shown in Fig.~\ref{fig:scales}  with the highest reach necessarily in $uc$-transitions.
Bounds on gluonic dipole operators are the strongest,  $ \SI{11.4}{\TeV}$,  followed by the EW dipole operators $ \SI{5.6}{\TeV}$, the four-fermion operators $ \SI{3.6}{\TeV}$, and  the $Z$-penguin operators $ \SI{1.6}{\TeV}$, consistent with  the shapes and enhancements discussed in Sec.~\ref{sec:shapes}.
  In particular, the $Z$-penguins are not energy-enhanced. The limits from LEP on these operators are therefore about an order of magnitude stronger 
\cite{Efrati:2015eaa,Bellafronte:2023amz}  than the ones obtained in our analysis. 

\begin{figure}[h]
  \centering
  \includegraphics[width =0.8 \linewidth]{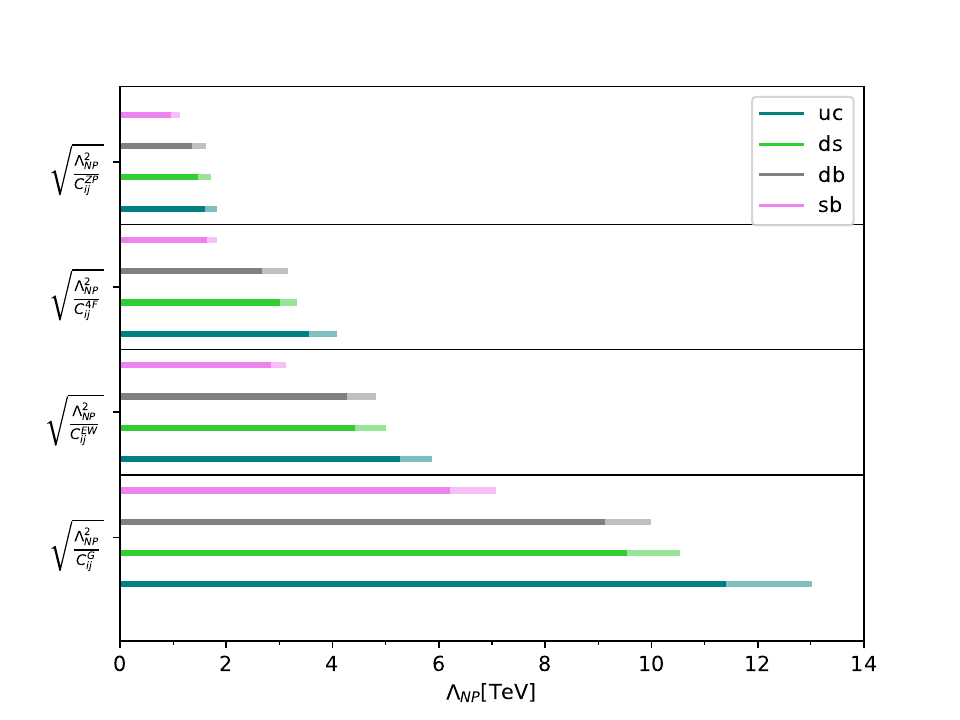}
  \caption{NP scales probed  in  $pp \to \text{MET} +j$ via FCNC transitions corresponding to the  limits in  Table \ref{tab:fcnc_bounds}.
  The vertical axis shows,  from top to bottom and with increasing relative reach, the limits from $Z$-penguin, four-fermion, electroweak and gluon dipole  insertions
  for the quark flavor changes  $sb$ (pink), $db$ (brown), $ds$ (green) and $uc$ (dark green). }
  \label{fig:scales}
\end{figure}

In  Table \ref{tab:dipole_bounds} we   present limits on  flavor conserving dipole operators. Note, their  interference with the SM is chirality suppressed, i.e., involves a fermion mass,
which we safely neglect. 
\begin{table}[h]
  \centering
  \begin{tabular}{c|c|c}
  \toprule
  $ii$  &   $C^{G}_{ii}/\sqrt{2}$ &  $C^{EW}_{ii}/\sqrt{2}$\\
  \midrule
   $uu $                & \num{0.0051}(\num{0.0039})  &   \num{0.023}(\num{0.018})  \\
   $dd $                  & \num{0.0079}(\num{0.0062}) &   \num{0.034}(\num{0.027})  \\
   $ss $                  & \num{0.023}(\num{0.018}) &   \num{0.091}(\num{0.075})  \\
   $cc $                & \num{0.028}(\num{0.023}) &    \num{0.11}(\num{0.09}) \\
   $bb $                  & \num{0.038}(\num{0.031})&    \num{0.17}(\num{0.14})  \\ 
  \bottomrule
  \end{tabular} 
  \caption{ The $95\%$ CL-limits on effective  gluon dipole WCs \eqref{eqn:gluon_Eff} and  electroweak $Z$-dipole WCs \eqref{eqn:EW_dip} quark flavor conserving combinations, see Table \ref{tab:fcnc_bounds}. The factor  $1/\sqrt{2}$ accounts for the fact that the Wilson coefficients for the chirality-flipped insertions coincide for diagonal flavors. } 
  \label{tab:dipole_bounds}  
\end{table}
The bounds  for the  diagonal couplings  also follow the luminosities, that is, $uu$ has  the strongest bound,  probing scales up to 14 TeV,
followed by $dd$, $ss$, $cc$ and $bb$. 

Increasing the luminosity to $\SI{3000}{\femto\barn}^{-1}$  would tighten the limits  in Tables \ref{tab:fcnc_bounds}, \ref{tab:dipole_bounds}  by  $(70-80) \%$.

\begin{figure}
  \centering
  \includegraphics[width = 0.48\linewidth]{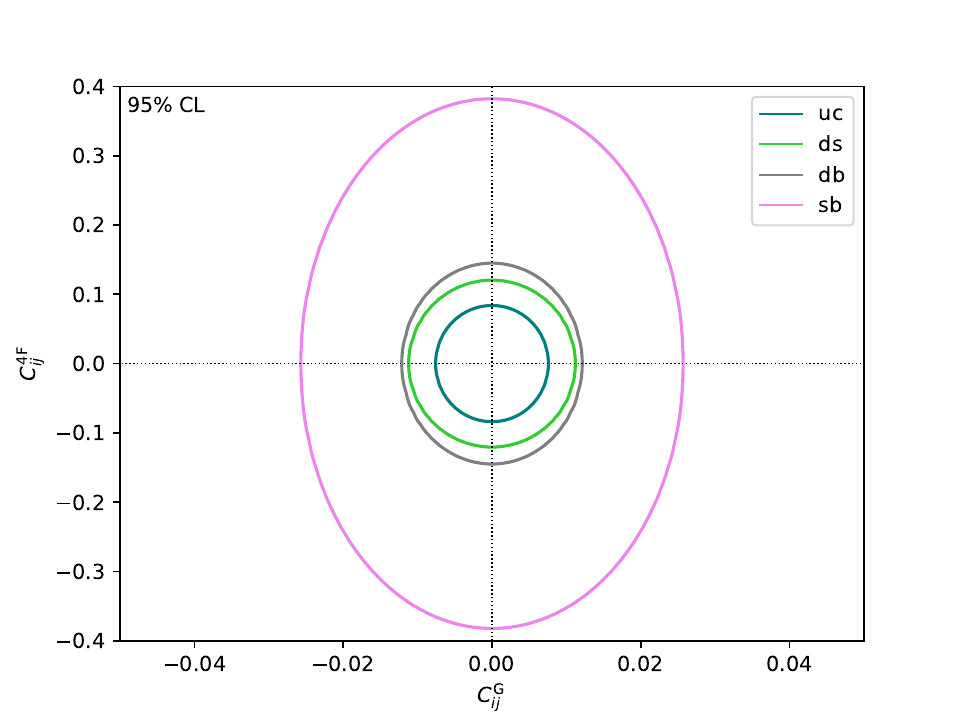}
   \includegraphics[width =0.48 \linewidth]{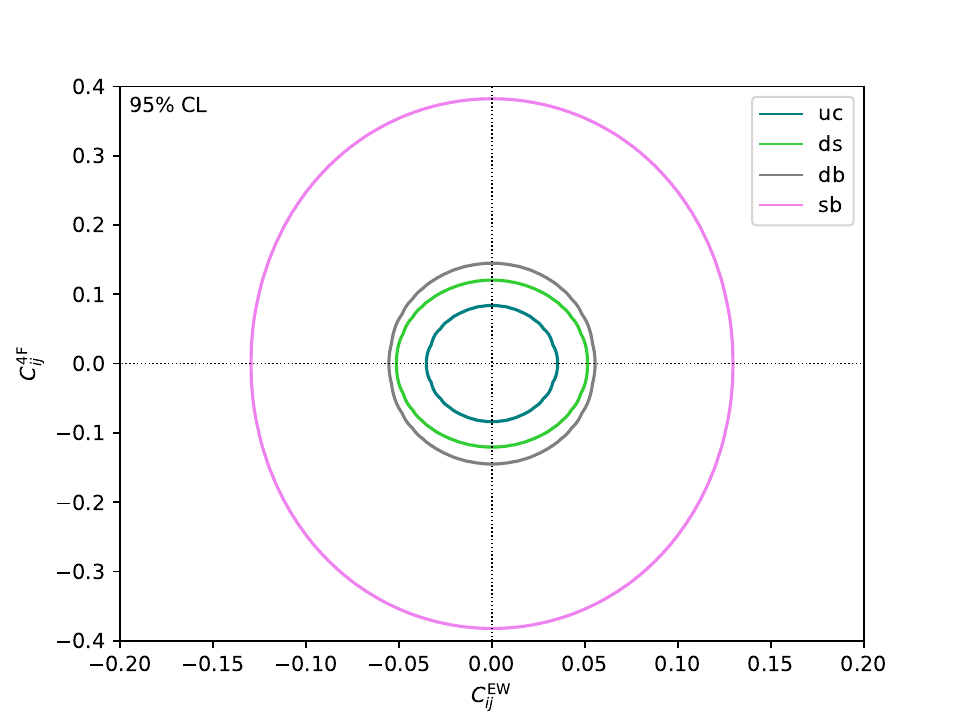}
    \includegraphics[width = 0.48\linewidth]{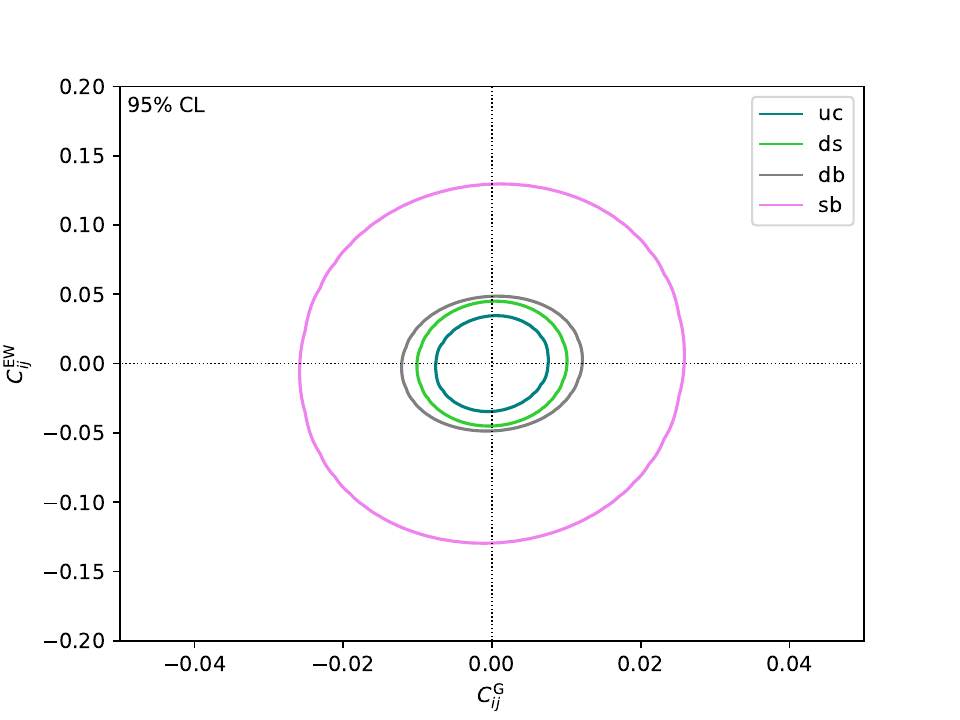}
  \caption{Two dimensional correlations of 4-Fermion - gluon dipoles (upper left),  4-Fermion - EW dipole operators  (upper right), and EW - gluon dipole operators  (lower plot) for the quark flavor changes
   $sb$ (pink), $db$ (brown), $ds$ (green) and $uc$ (dark green). }
   \label{fig:2D}
\end{figure}
We also consider 2D-correlations  in Fig.~\ref{fig:2D} showing  operator  combinations involving the coefficients that are most strongly constrained. 
Operators do not interfere in  the  cross sections (\ref{eq:sig}) except for the  EW and gluon dipoles (lower plot), however this term is very small.

\subsection{Limits on light right-handed neutrinos \label{sec:RHN}}

Light right-handed neutrinos $N$, which are singlets of the SM, can be added to the dimension-six SMEFT-lagrangian as
\begin{equation}
    \label{eqn:lagrange_N}
    \begin{aligned}
      \mathcal{L}^{6}_{\text{SMEFT}+N} &\supset \frac{ C_{Nq,kl ij }}{\Lambda_{NP}^2}\left(\bar N_k \gamma_{\mu}  N_l \right)\left( \bar q_i \gamma^{\mu}  q_j \right)             + \frac{C_{Nu,kl ij }}{\Lambda_{NP}^2}\left(\bar N_k \gamma_{\mu}  N_l \right)\left( \bar u_i \gamma^{\mu}  u_j \right) \\
      &+\frac{C_{Nd,kl ij }}{\Lambda_{NP}^2}\left(\bar N_k \gamma_{\mu}  N_l \right)\left( \bar d_i \gamma^{\mu}  d_j \right) 
    \end{aligned}
  \end{equation}
  where $k,l$ are flavor indices of the $N$'s. In principle, the numbers of flavors for the singlet neutrinos can be different from the left-handed ones.
Feynman diagrams contributing to $pp \to \text{MET} +j$ are given  in Fig.~\ref{fig:parton_feyn} with the $\nu$'s replaced by the $N$'s. 
Assuming  $Z$-decays to invisibles to be SM-like, there is no contribution from the singlet neutrinos via  dipole or penguins operators.
Note that further (pseudo-)scalar four-fermion operators with both a left-handed and a right-handed neutrino exist 
but these are not considered in this work.
The Wilson coefficients for the singlets
$C_{Nq}, C_{Nu}, C_{Nd}$ can be constrained just like  the ones involving left-handed neutrinos  using  (\ref{eqn:WC_eff}) as 
\begin{equation}
    \label{eqn:WC_eff-N}
    C^{4F+N}_{ij} =\sqrt{  |C^{4F}_{ij}|^2+  \sum_{k,l}  | C_{Nq, kl ij}|^2 + |C_{N u/d ,kl ij}|^2} \, . 
\end{equation}
Therefore, upper limits on   $C^{4F}_{ij}$ given  in  Table \ref{tab:fcnc_bounds} apply to $ C^{4F+N}_{ij}$.

\subsection{Comparison of bounds on four-fermion operators \label{sec:comp}}

We compare  the  limits on four-fermion operators $C_{ij}^{4F}$  from $pp \to \text{MET}+j$ given in   Table \ref{tab:fcnc_bounds}  to existing  ones from Drell-Yan  production of charged leptons and rare decays  of kaons,  charm and beauty mesons.
We make use of  $SU(2)_L$-links,  that  enable a representation of dineutrino limits  in terms of left-handed dilepton coefficients 
$\mathcal{K}_{L,R}^{q q^\prime \ell \ell^\prime} $, where  the subscript  indicates  left ($L$), right $(R$)-chiral quark $|\Delta q|=|\Delta q^\prime|=1$ FCNCs.
Limits on up-type FCNCs with dineutrinos  imply  limits on right-handed up-type currents  and left-handed down-type currents with charged leptons,
and vice versa.
We follow  closely  \cite{Bause:2020auq}, and give 
details on the SMEFT-WET matching and $SU(2)_L$-machinery  in App.~\ref{app:WET_FF}.
Coefficients for lepton flavor violating (LFV) transitions are given as charge-summed
\begin{equation*}
  \overline{ \mathcal{K}^{\ell+\ell'-}_{L,R}}=  \sqrt{|\mathcal{K}_{L,R}^{\ell+\ell'- }|^2 + |\mathcal{K}_{L,R}^{\ell-\ell'+} |^2 } \, . 
\end{equation*}

Constraints on four-fermion operators  are summarized  in Tables \ref{tab:WET_uc}, \ref{tab:WET_sd}, \ref{tab:WET_bd} and \ref{tab:WET_bs} for $uc$, $ds$, $db$ and $sb$, respectively.
Top quark couplings can as well  be constrained  by  $SU(2)_L$ and $b$-physics data,  resulting limits are shown  in Tables \ref{tab:WET_tu} and \ref{tab:WET_tc} for $tu$ and $tc$ couplings, respectively. Note, the tables are  in parts adopted from \cite{Bause:2020auq}, with updates from \cite{Grunwald:2023nli}.
The novel  entries obtained in this work are  in the last rows, from  the process $pp \to \nu \bar \nu +X$. We discuss the impact on the  individual sectors below.

\subsubsection{ $cu\ell \ell^\prime$-transitions}
\noindent

The best limits on $\mathcal{K}^{cu\ell\ell'}_L$ are from kaon decays, $K \to \pi \nu \bar \nu$, about two orders of magnitude stronger than the other $cu \ell \ell^\prime$-ones.
The best limits on $\mathcal{K}^{cu\ell\ell'}_R$ are from  $pp \to \nu \bar \nu +X$ if taus are involved: $\ell \ell^\prime=\tau \tau, e \tau, \mu \tau$.
For the other lepton flavors they are competitive with Drell-Yan and  rare $D$-decays, where the latter dominates  the $\mathcal{K}^{cu\mu\mu}_R$- limit.

\begin{table*}[h]
  \centering
  \begin{tabular}{lc |c c c c c c c}
  \toprule
    Process & WC& $ee$ & $\mu \mu$ & $\tau \tau $& $e \mu$ & $e \tau $ &$\mu \tau $ \\
    \midrule 
    $pp \rightarrow \ell^+\ell^- \qquad$           &$\mathcal{K}^{cu\ell\ell'}_{L,R}\qquad$& 2.9 & 1.6 & 5.6& 1.6 & 4.7 &5.1 \\
    $D \rightarrow  \ell^+\ell^- + \pi $              &$\mathcal{K}^{cu\ell\ell'}_{L,R}\qquad$& 4.0 & 0.9 & -  & 2.2 & -   & -  \\
    $K \rightarrow \nu \bar \nu \quad  + \pi$         &$\mathcal{K}^{cu\ell\ell'}_{L}\cdot \num{e2}$&$ [-1.9,0.7]$& $ [-1.9,0.7]$ & $ [-1.9,0.7]$  & \num{1.1} & \num{1.1}   &\num{1.1}  \\
    $pp \rightarrow \nu \bar \nu \quad + X$     &$\mathcal{K}^{cu\ell\ell'}_{L}\qquad$& 5.7 (4.6)  & 5.7 (4.6) & 5.7  (4.6) & 4.1 (3.3)  &  4.1 (3.3) & 4.1 (3.3) \\
    $pp \rightarrow \nu \bar \nu \quad + X$     &$\mathcal{K}^{cu\ell\ell'}_{R}\qquad$& 4.1 (3.3)  & 4.1 (3.3)  & 4.1 (3.3)  &2.9 (2.2)  &  2.9 (2.2) & 2.9 (2.2) \\
    \bottomrule
  \end{tabular}
  \caption{  Limits on  $cu\ell \ell^\prime$ WCs. The first three rows are taken from \cite{Bause:2020auq}, while the 
  ones from $pp  \to  \nu \bar \nu + X$  are based on table  \ref{tab:fcnc_bounds}, with  projections for $\SI{3000}{ \femto \barn}^{-1}$ in parentheses, using $SU(2)$~ \cite{Bause:2020auq}.  LFV-bounds are given as flavor-summed. }
  \label{tab:WET_uc}
\end{table*}

\subsubsection{$sd\ell \ell^\prime$-transitions}

The best limits on $\mathcal{K}^{sd\ell\ell'}_R$  (from $K \to \pi \nu \bar \nu$) and 
$\mathcal{K}^{sd\ell\ell'}_L$  for $\ell \ell^\prime = ee, \mu \mu, e \mu$, are from rare $K$-decays, about two orders of magnitude stronger than all  others.
The best limits on $\mathcal{K}^{sd\ell\ell'}_L$  with taus, $\ell \ell^\prime=\tau \tau, e \tau, \mu \tau$ are from  $pp \to \nu \bar \nu +X$.
Dilepton Drell-Yan (first row) and $\text{MET} +j$ limits (last two rows) are of similar size.
\begin{table*}[h]
    \centering
    \small
    \begin{tabular}{lc |c c c c c c c}
      \toprule
      Process &WC & $ee$ & $\mu \mu$ & $\tau \tau $& $e \mu$ & $e \tau $ &$\mu \tau $ \\
      \midrule 
      $pp \rightarrow \ell^+\ell^- \qquad$               &$\mathcal{K}^{sd\ell\ell'}_{L,R}\qquad $ & 3.8 & 2.3 & 5.37& 2.0 & 6.1 &6.6 \\
      $K \rightarrow  \ell^+\ell^- + \pi$                  &$\mathcal{K}^{sd\ell\ell'}_{L,R}\cdot \num{e2}  $  &\num{5} & \num{1.6} & -  & \num{6.6e-2} & -   & - \\
      $K \rightarrow \nu \bar \nu \quad + \pi$           &$\mathcal{K}^{sd\ell\ell'}_R\cdot \num{e2}   $ & [-1.9,0.7]& [-1.9,0.7] & [-1.9,0.7]  & \num{1.1} & \num{1.1}   &\num{1.1}   \\
      $pp \rightarrow \nu \bar \nu \quad + X$   &$\mathcal{K}^{sd\ell\ell'}_{L}\qquad$  &4.1 (3.3)  & 4.1 (3.3)  & 4.1 (3.3)  &2.9 (2.2)  &  2.9 (2.2) & 2.9 (2.2) \\
      $pp \rightarrow \nu \bar \nu \quad + X$   &$\mathcal{K}^{sd\ell\ell'}_{R}\qquad$  &5.7 (4.6)  & 5.7 (4.6) & 5.7  (4.6) & 4.1 (3.3)  &  4.1 (3.3) & 4.1 (3.3) \\
      \bottomrule
    \end{tabular}
    \caption{Limits on $sd\ell \ell^\prime$ WCs. The first row is based on \cite{Grunwald:2023nli}, while the second and third row are taken from \cite{Bause:2020auq}.
    See table \ref{tab:WET_uc}.}
    \label{tab:WET_sd}
\end{table*}

\subsubsection{$bq\ell \ell^\prime$-transitions}

For WCs involving $b$-quarks, the  $pp \rightarrow \nu \bar \nu + X$ analysis  (last row of tables  \ref{tab:WET_bd} and \ref{tab:WET_bs})  can only provide  constraints on right-chiral WCs $\mathcal{K}^{bq\ell\ell'}_{R}$. For both $q=d,s$ FCNCs the corresponding limits from  rare $B$-meson decays into invisibles (second to last rows)
are stronger, and more so for $b \to s$, because the PLF is smaller and, on the other hand, the $ b \to s \nu \bar \nu$ branching ratios are larger than the ones for $b \to d$.
When taus are not involved, precision studies of  semileptonic rare $B$-decays give more than an   order of magnitude better bounds 
than from high $p_T$-studies.

\begin{table*}[h]
  \centering
  \begin{tabular}{lc |c c c c c c c}
    \toprule
    Process & WC & $ee$ & $\mu \mu$ & $\tau \tau $& $e \mu$ & $e \tau $ &$\mu \tau $ \\
        \midrule 
    $pp \rightarrow \ell^+\ell^- \qquad$                       &$\mathcal{K}^{bd\ell\ell'}_{L,R} $           & 5.4 & 3.2 & 6.5& 3.1 & 9.6 &11 \\
    $B \rightarrow  \ell^+\ell^- + X$                           &$\mathcal{K}^{bd\ell\ell'}_{R} $                  &0.09 & [-0.03, 0.03] & 21  & \num{0.2} & 3.4   & 2.4  \\
    $B \rightarrow  \ell^+\ell^- + X$                           &$\mathcal{K}^{bd\ell\ell'}_{L} $                  &0.09 & [-0.07,-0.02] & 21  & \num{0.2} & 3.4   & 2.4  \\
    $B \rightarrow \nu \bar \nu +(\pi,\rho)$            &$\mathcal{K}^{bd\ell\ell'}_{R} $ & 1.8&1.8 & 1.8  & 2.5 & 2.5  &2.5  \\
    $pp \rightarrow \nu \bar \nu \quad + X$     &$\mathcal{K}^{bd\ell\ell'}_{R}$   & 7.3(5.2)  & 7.3(5.2)   &7.3(5.2)   &5.2(3.7)   &  5.2(3.7)  & 5.2(3.7)  \\
    \bottomrule
  \end{tabular}
  \caption{Constraints on  $bd\ell \ell^\prime$ WCs. The first row is based on \cite{Grunwald:2023nli}, while the second, third and fourth row are taken from \cite{Bause:2020auq}. 
     See table \ref{tab:WET_uc}.}
  \label{tab:WET_bd}
\end{table*}

\begin{table*}[h]
  \centering
  \begin{tabular}{ lc |c c c c c c c}
    \toprule
    Process &WC & $ee$ & $\mu \mu$ & $\tau \tau $& $e \mu$ & $e \tau $ &$\mu \tau $ \\
      \midrule 
    $pp \rightarrow  \ell^+\ell^- \qquad$                 &$\mathcal{K}^{bs\ell\ell'}_{L,R}$ & 15 & 8.9 & 17& 8.0 & 27 &30 \\
    $B \rightarrow \ell^+\ell^- + X$                    &$\mathcal{K}^{bs\ell\ell'}_{R} $  &0.04 & [-0.03,-0.01] & 32  & \num{0.1} & 2.8   & 3.4  \\
    $B \rightarrow \ell^+\ell^- + X$                    &$\mathcal{K}^{bs\ell\ell'}_{L} $&0.04 & [-0.07,-0.04] & 32  & \num{0.1} & 2.8   & 3.4  \\
    $B \rightarrow \nu \bar \nu   \quad +K^{(*)}$           &$\mathcal{K}^{bs\ell\ell'}_{R}  $ & 1.4&1.4 & 1.4  & 1.8 & 1.8   &1.8  \\
    $pp \rightarrow \nu \bar \nu \quad+ X$      &$\mathcal{K}^{bs\ell\ell'}_{R}$  & 19.3(15.3)  & 19.3(15.3)   &19.3(15.3)   &13.6(11.1)   &  13.6(11.1)  & 13.6(11.1)  \\
    \bottomrule
  \end{tabular}
  \caption{Limits on $bs\ell \ell^\prime$ WCs. The first row is based on \cite{Grunwald:2023nli}, while the second, third and fourth row are taken from \cite{Bause:2020auq}. 
   See table \ref{tab:WET_uc}.}
  \label{tab:WET_bs}
\end{table*}

\subsubsection{$tq\ell \ell^\prime$-transitions}

For  FCNCs involving $t$-quarks, the  $pp \rightarrow \nu \bar \nu + X$ analysis (last row)  can only  constrain left-chiral WCs
$\mathcal{K}^{tq\ell \ell^\prime}_L$ (last rows tables \ref{tab:WET_tu} and \ref{tab:WET_tc}). For both $ut$ and $ct$ FCNCs the corresponding $SU(2)$-limits  from rare $B$-meson decays into invisibles (second rows)
are stronger, and more so for $t \to c$ than for $t \to u$ due to effects already discussed for $bq\ell \ell^\prime$-transitions.
To derive limits on $\mathcal{K}^{tq\ell \ell^\prime}_R$  one would need  top-physics  data, which  would also constrain left-chiral $b$-quark couplings. 

\begin{table*}[h]
    \centering
    \begin{tabular}{lc |c c c c c c c}
      \toprule
      Process &WC & $ee $& $\mu \mu$ & $\tau \tau $& $e \mu$ & $e \tau $ &$\mu \tau $ \\
          \midrule 
      $t \: \:\rightarrow \ell^+ \ell^- u$                  & $\mathcal{K}^{tu\ell \ell'}_{L,R}$           & $\sim  200$ & $\sim 200$ & n.a.& 12 & 136 &136 \\
      $B \:\rightarrow \nu \bar \nu + (\pi,\rho)$                 & $\mathcal{K}^{tu\ell \ell '}_L    $  & [-1.6,1.8]&[-1.6,1.8] & [-1.6,1.8]  & 2.4 & 2.4  &2.4  \\
      $pp \rightarrow \nu \bar \nu + X$ & $\mathcal{K}^{tu\ell \ell'}_{L}  $  & 7.3(5.2)  & 7.3(5.2)   &7.3(5.2)   &5.2(3.7)   &  5.2(3.7)  & 5.2(3.7) \\
      \bottomrule
    \end{tabular}
    \caption{Limits on $tu\ell \ell^\prime$ WCs. The first two rows are taken from \cite{Bause:2020auq}.
     See table \ref{tab:WET_uc}.}
    \label{tab:WET_tu}
\end{table*}
\begin{table*}[h]
    \centering
    \begin{tabular}{ lc |c c c c c c c}
      \toprule
      Process&WC& ee & $\mu \mu$ & $\tau \tau $& $e \mu$ & $e \tau $ &$\mu \tau $ \\
    \midrule 
      $t \rightarrow \ell^+ \ell^- c$                  &   $\mathcal{K}^{tc\ell \ell'}_{L,R} $           & $\sim 200$ & $\sim 200$ & n.a.& 36 & 136 &136 \\
      $B \rightarrow \nu \bar \nu + K^{(*)}$                 & $\mathcal{K}^{tc\ell \ell'}_L   $ & [-1.9,0.9]&[-1.9,0.9] & [-1.9,0.9]  & 1.8 & 1.8  &1.8  \\
      $pp \rightarrow \nu \bar \nu \:\:\:+ X$ &   $\mathcal{K}^{tc\ell \ell'}_{L}$  & 19.3(15.3)  & 19.3(15.3)   &19.3(15.3)   &13.6(11.1)   &  13.6(11.1)  & 13.6(11.1)   \\
      \bottomrule
    \end{tabular}
    \caption{Limits on  $tc\ell \ell^\prime$ WCs. The first two rows are taken from \cite{Bause:2020auq}.
     See table \ref{tab:WET_uc}.}
    \label{tab:WET_tc}
\end{table*}

\subsubsection{Synopsis four-fermion bounds}

The constraints derived in this work from  $pp \rightarrow \nu \bar \nu +  X$ give the presently strongest limits on  left-handed $|\Delta s|=|\Delta d|=1$ FCNCs and right-handed $|\Delta c|=|\Delta u|=1$ FCNCs in couplings with  taus  involved, that is,  for $\ell \ell^\prime=\tau \tau, e \tau, \mu \tau$ operators.  All other cases in the first two generations of quarks are dominated by bounds from rare kaon decays, see  Tables \ref{tab:WET_uc}, \ref{tab:WET_sd}. 
For other couplings, $pp \rightarrow \nu \bar \nu  + X$ limits are comparable to those from Drell-Yan or rare charm decays.
Dineutrino rare  $B$-decays, however, provide stronger constraints.
The novel limits are however also interesting as they are complementary and probe
various theoretical and experimental systematics. We look forward to improved analyses of MET plus jet.

\subsection{Comparison of  bounds on dipole operators \label{sec;dipole-comp}}

The  limits on the gluon dipole coefficients given in Table \ref{tab:dipole_bounds}  improve on previous collider studies,  here given for 
$\Lambda_{NP} = \SI{1}{\TeV}$ \cite{Haisch:2021hcg}
\begin{align}
C_{uG, cc}, C_{ dG, bb} \lesssim 7 ~~(\text{dijet angular distribution}) , \quad C_{dG, bb} \lesssim 0.40 ~~(\text{$b$-jets}) 
\end{align}
by  two orders of magnitude for $cc$ and one for $bb$.
For completeness, a fit to top-observables obtains constraints on the dipole couplings (for $\Lambda_{NP} = \SI{1}{\TeV}$) as  \cite{Grunwald:2023nli}
\begin{align}
C_{uG, tt} \lesssim 0.028, \quad C_{uB, tt} \lesssim 0.38, \quad C_{uW, tt} \lesssim 0.053 \, . 
\end{align}
 Limits on $ C_{qB, ij},  C_{qW, ij}$ have been obtained from dilepton tails  at the LHC \cite{Allwicher:2022gkm}, which are a bit weaker than the ones in
Tables \ref{tab:fcnc_bounds}, \ref{tab:dipole_bounds}.

For the discussion of constraints from low energy let us make some general remarks relevant for dipole operators.
Firstly, severe  constraints exist  from CP-violating observables, notably electric dipole moments,  but also CP-asymmetries, e.g. \cite{Haisch:2021hcg,Fajfer:2023gie}
on the imaginary parts of the dipole coefficients. The $pp \to \nu \bar \nu +X$  study, on the other hand, probes their modulus, hence is complementary.
Secondly, our analysis probes the $Z$-direction  of the electroweak coefficients $C_{u/d B}$ and $C_{u/d W}$  (\ref{eqn:EW_dip}), while in low energy observables
the photon-one is constrained. It is  therefore useful to study in the future also $pp \to \nu \bar \nu + \gamma$, to access the photon-coupling directly, and to disentangle the
hypercharge and $SU(2)_L$ contributions.
In the following, we  perform the comparison of constraints with only the gluon dipole operator $O_{u/dG}$ present at the high scale.

The dipole coefficients $C_{u/dX}$ ($X = G,W,B$) are subject  to sizable effects from renormalization group running and mixing within SMEFT \cite{Alonso:2013hga}, and once matched onto the 
 low energy theory, continue to do so.
 The main contributors in the low energy effective theory are  the  electromagnetic  $C_7^{(\prime)}$ and gluonic $C_8^{(\prime)}$ dipole coefficients,
see  App~\ref{app:WET_dipole}. 
As argued, from $pp \to \nu \bar \nu +X$  we do not probe the photon coupling and hence neglect contributions from the electroweak dipole operators
at the high scale. On the other hand,  $C_7^{(\prime)}$ is induced by running and mixing with the gluon dipole operator.

For the numerical evaluation of matching and running we employ the package \texttt{wilson} \cite{wilson}.
The input is provided by the SMEFT WCs Table  \ref{tab:fcnc_bounds}  at $\Lambda_{NP} = \SI{1}{\TeV}$.
These bounds are  evolved  to the electroweak scale  and matched onto the WET.
The limits  are further evolved within WET,  down to the $b$-mass scale $\approx m_b$, where the results for $bs$ and $bd$ transitions are extracted. 
For charm-transitions, the $b$-quark is integrated out and the running is performed until the charm scale $\approx m_{c}$, where the bounds for $uc$-transitions are extracted. 
Quark masses are  MS-bar masses at the corresponding mass scale.
For $ds$-transitions, the charm-quark is integrated out and the running is performed until $ \sim \SI{1}{\GeV}$.
The constraints  on $C_7,C'_7,C_8,C'_8$ are presented in Table \ref{tab:dipole_WET}.

\begin{table}[h]
  \centering
  \begin{tabular}{c| c| c c c c}
  \toprule
  $qq'$  & $\mu$ & $|C_7|$ & $|C_7'|$ & $|C_8|$ & $|C_8' |$  \\
  \midrule
   $uc$      & $m_c$        &  \num{2.8}     &\num{2.8}      & \num{4.4} & \num{4.4}\\ 
   $ds$       & 1 GeV       &  \num{7.1e4} & \num{7.1e4} & \num{2.2e5}        &\num{2.2e5}  \\
   $db$        & $m_b$         &  \num{240}    &\num{240} & \num{48} & \num{49} \\
   $sb$        & $m_b$         &  \num{110}   &\num{120}      & \num{23} & \num{23}  \\
  \bottomrule
  \end{tabular} 
  \caption{The $95 \%$ CL upper limits on  the WET coefficients at the scale $\mu$
     from  running and matching the gluonic dipole limits in SMEFT Table  \ref{tab:fcnc_bounds}, see text.
  The  photon dipole coefficients $C_7^{(\prime)}$ are  induced by mixing with the gluon dipole operators.} 
  \label{tab:dipole_WET}
\end{table}

Since the leading effect is QCD-running and QCD respects parity, the left-handed and right-handed sectors evolve independent from each other, and
a splitting between $C_{7,8}$ and the chirality-flipped ones $C_{7,8}^\prime$ are only induced from SM-contributions and subdominant.

Due to the strong GIM-suppression the situation in charm is simpler than in the down-sector and the approximate formula is useful \cite{deBoer:2017que,Adolph:2020ema}
\begin{align}  \label{eq:Cuc}
C_7^{(\prime)}(m_c) \simeq 0.4 \big( C_7^{(\prime)}(\Lambda_{NP}) -  C_8^{(\prime)}(\Lambda_{NP}) \big) \,, 
\qquad C_8^{(\prime)}(m_c) \simeq  0.4 C_8^{(\prime)}(\Lambda_{NP})   \,.
\end{align}
It is valid to roughly  20\%  for  $\Lambda_{NP}$ 
within 1-10 TeV,  and for $C_7^{(\prime)}(\Lambda_{NP}) =0$  consistent with Table \ref{tab:dipole_WET} which is based on \texttt{wilson}.
The branching ratios of $D^0 \to \rho^0 \gamma$  and $D \to \pi \mu^+ \mu^-$ constrain $| C_7^{(\prime)}(m_c)| \lesssim 0.3$ \cite{deBoer:2017que,Golz:2021imq}.
Confronting this to  the upper limit  in Table \ref{tab:dipole_WET} which  stems from the gluon dipole constraint, we learn from
Eq.~(\ref{eq:Cuc}) that there has to be a cancellation tuned to one order of magntitude between the photonic and gluonic contributions at high energies or
the limits from $pp \to \nu \bar \nu +X$ on the gluon dipole operators are about one order of magnitude weaker than the ones from rare charm decays.
This highlights the importance of collider studies including photons such as $ p p \rightarrow  \gamma + \nu \bar \nu$, which directly contributes to $C_7^{(\prime)}$. 
Constraints on the imaginary parts from low energy data are  even stronger, 
 $|\text{Im}(C_8^{(\prime)}(m_c))| \lesssim {\mathcal{O}}(0.1)$  from the CP-asymmetry  $A_{CP}(D^0 \rightarrow \rho^0 \gamma)$
\cite{deBoer:2017que,Lyon:2012fk},
 and  $|\text{Im}(C_8^{(\prime)}(m_c))| \lesssim \num{e-3}$  from CP-violation in hadronic 2-body $D^0$-decays, $\Delta A_{CP}$ \cite{deBoer:2017que,Giudice:2012qq}.

The limits from global fits of $B$-decay data   on  the gluon dipole coefficients  are  more than one order of magnitude stronger
than the  ones  from $ pp \rightarrow \nu \bar \nu + X$,
$| C_8^{(\prime)} (m_b) | \lesssim 1.6$  for $b \rightarrow s$ and $| C_8 (m_b) | \lesssim 3$   for $b \rightarrow d$ transitions.
The corresponding constraints on the photon dipole are even stronger for $b \rightarrow s$ modes $| C_7^{(\prime)} (m_b)| \lesssim 0.1$ 
and for $b \rightarrow d $-transitions $| C_7(m_b) | \lesssim 2$ \cite{Bause:2022rrs,Mahmoudi:2023upg}.
Again the bounds from Table \ref{tab:dipole_WET} are several orders of magnitude weaker and are therefore not competitive.

For $s \rightarrow d $-transitions no comparable bounds on the modulus of the dipole couplings are found in the literature, but they are the least stringent ones in Table \ref{tab:dipole_WET} and therefore are also not expected to be competitive.

\section{Summary \label{sec:sum}}

We work out constraints on new physics in the SMEFT  by recasting an ATLAS search \cite{ATLAS:2021kxv} for large missing transverse energy and an energetic jet at the LHC.
The limits are presented in Tables \ref{tab:fcnc_bounds}, \ref{tab:dipole_bounds}, and given in  terms of effective WCs
of semileptonic four-fermion operators, gluon and electroweak dipole operators and $Z$-penguins.
Having all types of operators simultaneously present such as in a global fit would result in ellipsoids, with hierarchical
semi-axes, as in the 2D -examples in Fig.~\ref{fig:2D}, for any quark FCNC. Fitting different  quark FCNCs simultaneously requires a model of flavor, for example  MFV as in  \cite{Grunwald:2023nli}.
The effective  coefficients \eqref{eqn:gluon_Eff},  \eqref{eqn:EW_dip}, \eqref{eqn:WC_eff},  and \eqref{eqn:WC_eff_penguin} include flat directions that cannot be resolved using dineutrino plus jet data alone. We hope to come back in the future to combine this
in a more global analysis.

Performing  an explicit analytical computation of partonic cross sections we  show that all operators except the  $Z$-penguins  are 
energy-enhanced, see Sec.~\ref{sec:shapes}.
The gluon dipole operators have the highest sensitivity to NP and  probe energy scales   up to $14$ TeV (the $uu$-coupling). 
We also obtain significantly improved limits on $C_{uG,cc}$ and $C_{dG,bb}$ over  previous collider ones.
To disentangle the hypercharge from the $SU(2)_L$ contributions to the $Z$-dipole coefficients (\ref{eqn:EW_dip}})  study of 
$p p \rightarrow \nu \bar \nu + \gamma$ \cite{ATLAS:2020uiq} is encouraged, which directly probes the orthogonal, radiative dipole operator.
This would also benefit synergies between high-$P_T$ and low energy observables, as there is large mixing among all dipole operators.

The constraints on four-fermion operators from  $pp \to \text{MET}+j$  are presently the strongest  for left-handed $|\Delta s|=|\Delta d|=1$ FCNCs and right-handed $|\Delta c|=|\Delta u|=1$ FCNCs
if taus are involved, that is,  for $\ell \ell^\prime=\tau \tau, e \tau, \mu \tau$ operators. All other cases in the first-second generation of quarks are dominated by bounds from rare kaon decays, see  Tables \ref{tab:WET_uc}, \ref{tab:WET_sd}. 
The bounds from $pp \rightarrow \nu \bar \nu + X$ are  comparable to those from conventional Drell-Yan production,  except 
for tau flavors for which  the MET-ones are better. This can be understood by noting that  in $pp \rightarrow \ell^+ \ell^{\prime -}$ analyses  the 
 $\tau$-tagging is typically  inferior to $e$- or $\mu$-tagging.
 We also stress that the extra jet allows for contributions from  initial state gluons, which have larger PDFs.

We also consider  light right-handed neutrinos as contributors  to missing energy, see Sec.~\ref{sec:RHN}.
Further  invisible light states can be probed in $pp \to \text{MET}+j$  which  could  complement more model-specific searches for dark sectors.
Present limits from low energy precision studies in $K \to \pi \nu \bar \nu$ and $B \to K^{(*)} \nu \bar \nu$ and $B \to (\pi,\rho)  \nu \bar \nu$
are superior to the current $\text{MET} +j$ sensitivity. Limits on rare charm decays to invisibles are presently not competitive with other constraints~\cite{Bause:2020xzj}, 
and  would require an  order of magnitude
improvement of the limit on the  $D \to \pi \nu \bar \nu$ branching ratio \cite{BESIII:2021slf}.

In conclusion, the  interpretation of  $pp \to \text{MET}+j$  provides new constraints, which are complementary to other searches from high and low energies, and
allow for a wider range of new physics  explorations.  We look forward to further studies with invisibles.

{\bf Note added: }During the finalization of this works an experimental study of cross sections for missing transverse energy and jets  by ATLAS appeared \cite{ATLAS:2024vqf}.
It is based on $\mathcal{L}_{int}= \SI{140}{\femto\barn}^{-1}$,  slightly larger than  the one we use for the SMEFT analysis \cite{ATLAS:2021kxv}.
We checked that SMEFT-limits derived from  \cite{ATLAS:2024vqf} are comparable to our results in  Tables \ref{tab:fcnc_bounds}, \ref{tab:dipole_bounds}, although
 weaker by a factor $\sim2$  despite the larger $P_T$-bin in \cite{ATLAS:2024vqf}. This is consistent since background uncertainties are larger in the latter
and also the sensitivity to a dark matter model is  lower than in  \cite{ATLAS:2021kxv}.

\acknowledgements

We are happy to thank Hector Gisbert-Mullor, Lara Nollen, Emmanuel Stamou and Mustafa Tabet for useful discussions.
G.H. would like to thank the CERN Theory Department  for kind  hospitality and support during the finalization of this work.

\appendix 
\section{Weak Effective Theory}
\label{app:WET}
Below the electroweak scale the $W$-boson and all particles heavier can be integrated out to obtain an EFT for low energy processes. 
This allows the interpretation of the high $P_T$-constraints, with others coming from low energy observables, such as decays.
The connection between SMEFT and WET is given by the matching at the EW scale $\mu_{EW} = \SI{160}{\GeV}$.
This matching procedure and the running between the different scales will be discussed in the following sections.
At first, the subsection\ref{app:WET_FF} focusses on semileptonic four-fermion operator, while the second subsection  \ref{app:WET_dipole} focusses on the matching as well as running of the dipole operators.

\subsection{Semileptonic four-fermion operators}
\label{app:WET_FF}

The effective Hamiltonians for four-fermion operators with two quarks ($i,j$) and two leptons ($k,l$) is given by
\begin{equation}
    \label{eqn:H_nunu}
    \mathcal{H}^{\nu \bar \nu}_{eff} = \frac{-4 G_F}{\sqrt{2}} \frac{\alpha_e}{4\pi} \sum_k \mathcal{C}_{k}^{P_{ij kl}} Q_n^{ij kl} + \text{h.c.} ,
  \end{equation}
for two dineutrinos and 
\begin{equation}
    \label{eqn:H_ll}
    \mathcal{H}^{\ell^- \ell^+}_{eff} = \frac{-4 G_F}{\sqrt{2}} \frac{\alpha_e}{4\pi} \sum_k \mathcal{K}_{k}^{P_{ij kl}} O_k^{ij kl} + \text{h.c.} ,
  \end{equation}
for charged leptons, see~\cite{Bause:2021cna,Bause:2020auq} to which we refer for details.
The superscript $P = D,U$ refers to the down or up sector, $\alpha_e$ denotes the fine structure constant and $G_F$ fermi's constant.
The calligraphic WCs are given in the mass basis and can each be written as the sum 
\begin{align}
    \label{eqn:NP}
    \mathcal{K}_{k}^{P_{ij kl}}  = \mathcal{K}_{k,SM}^{P_{ij kl}}  + \mathcal{K}_{k,NP}^{P_{ij kl}} \, ,  \\
    \mathcal{C}_{k}^{P_{ij kl}} = \mathcal{C}_{k,SM}^{P_{ij kl}} + \mathcal{C}_{k,NP}^{P_{ij kl}} \, , 
\end{align} 
where the SM contributions can be found in \cite{Bause:2021cna}.
The derived bound in the SMEFT is matched onto the WET and contribute only to the NP part of equation \eqref{eqn:NP}. 
Explicitly,
\begin{equation}
  \label{eqn:matching}
  \begin{aligned}
    C^{U_{ij kl}}_{L} &= K^{D_{ij kl}}_{L} =  \frac{2 \pi v^2}{\alpha_e\Lambda_{NP}^2}  \left( C^{(1)}_{lq_{kl ij}} + C^{(3)}_{lq_{kl ij}}  \right) \, ,  \\ 
  C^{D_{ij kl}}_{L} &= K^{U_{ij kl}}_{L} =  \frac{2 \pi v^2}{\alpha_e\Lambda_{NP}^2}  \left( C^{(1)}_{lq_{kl pr}} - C^{(3)}_{lq_{ijpr}}  \right) \, ,  \\
  C^{U_{ij kl}}_{R} &= K^{U_{ij kl}}_{R} =  \frac{2 \pi v^2}{\alpha_e\Lambda_{NP}^2} C_{lu_{kl ij}} \, ,  \\
  C^{D_{ij kl}}_{R} &= K^{D_{ij kl}}_{R} =  \frac{2 \pi v^2}{\alpha_e\Lambda_{NP}^2} C_{ld_{kl ij}} \, , 
  \end{aligned}
\end{equation}
where the upright coefficients denote the WET WCs in the gauge basis, which are related by a rotation to the calligraphic WCs in equation \eqref{eqn:H_ll} and \eqref{eqn:H_nunu}.
Note that the first equal sign connects charged dileptons couplings with the dineutrino couplings. These relations follow from the SU(2) structure of the SMEFT.
For the right-chiral singlets, the relations are straightforward. The left-chiral doublets get a sign change between $C^{(1)}_{lq}$ and $C^{(3)}_{lq}$, which connects the different quark sectors, i.e. $D \Leftrightarrow U$ as seen in the first two lines of \eqref{eqn:matching}. \\
The rotation to the mass basis can be done using $4$ rotation matrices.
In the quark sector there are two left-handed rotation matrices $V_{u,d}$ and two right-handed ones $U_{u,d}$,
whereas there are only two rotation matrices for the lepton sector, which are given by $V_l$ and $V_{\nu}$.
The rotations are then given by \cite{Bause:2021cna}
\begin{align}
  \label{eqn:coeff_rot}
  \mathcal{C}^D_L &= V^{\dagger}_{\nu} V_d^{\dagger} C^D_L V_d V_{\nu} \, ,  &  \mathcal{C}_R^D &= V_{\nu}^{\dagger} U_d^{\dagger} C_R^D U_d V_{\nu} \, , & \\
  \mathcal{K}^D_L &= V_l^{\dagger} V_d^{\dagger}      K_L^D V_d V_l    \, ,   & \mathcal{K}_R^D &= V^{\dagger}_l U^{\dagger}_d K_{R}^D U_d V_l. &
\end{align}
By summing over lepton flavors, expanding $V_{CKM} = V_u^{\dagger} V_{d}$ in $\lambda \sim 0.2$ and using the unitarity of the PMNS matrix $W = V^{\dagger}_l V_{\nu}$ leads to
 \cite{Bause:2020auq}
\begin{align}
    \label{eqn:SU2_links}
    \sum_{k,l = \nu_e,\nu_{\mu},\nu_{\tau}} \left( | \mathcal{C}^{ U_{ij kl} }_L |^2 + | \mathcal{C}^{U_{ij kl} }_R |^2 \right) &= \sum_{l,k = e,\mu,\tau} \left( | \mathcal{K}^{D_{ij kl}}_L |^2 + | \mathcal{K}^{U_{ij kl} }_R |^2 \right) + \mathcal{O}(\lambda)  \\
    \sum_{k,l = \nu_e,\nu_{\mu},\nu_{\tau}} \left( | \mathcal{C}^{ D_{ij kl} }_L |^2 + | \mathcal{C}^{D_{ij kl} }_R |^2 \right) &= \sum_{l,k = e,\mu,\tau} \left( | \mathcal{K}^{U_{ij kl}}_L |^2 + | \mathcal{K}^{D_{ij kl} }_R |^2 \right) + \mathcal{O}(\lambda) \, .
  \end{align}
Using equations \eqref{eqn:matching} and \eqref{eqn:SU2_links} and an explicit form for the effective WCs \eqref{eqn:WC_eff} can be found.
For up-type quarks this reads 
\begin{equation}
  \label{eqn:C_eff_up}
\begin{aligned}
    {C^{4F}_{ij}}^2 =&     \left(\frac{\alpha_e\Lambda_{NP}^2}{2 \pi v^2} \right)^2 \sum_{k,l = \nu_e,\nu_{\mu},\nu_{\tau}} |\mathcal{C}^{U_{ij kl}}_L |^2 + |\mathcal{C}^{U_{ij kl}}_R |^2  \\
               =&     \left(\frac{\alpha_e\Lambda_{NP}^2}{2 \pi v^2} \right)^2 \sum_{k,l = e,\mu,\tau} |\mathcal{K}^{D_{ij kl}}_L |^2 + |\mathcal{K}^{U_{ij kl}}_R |^2  + \mathcal{O}(\lambda) ,
  \end{aligned}
\end{equation}
  whereas for down-type quarks 
  \begin{equation}
  \begin{aligned}
    \label{eqn:C_eff_down}
    {C^{4F}_{ij}}^2 =&     \left(\frac{\alpha_e\Lambda_{NP}^2}{2 \pi v^2} \right)^2 \sum_{k,l = \nu_e,\nu_{\mu},\nu_{\tau}} |\mathcal{C}^{D_{ij kl}}_L |^2 + |\mathcal{C}^{D_{ij kl}}_R |^2 \\
                    =&     \left(\frac{\alpha_e\Lambda_{NP}^2}{2 \pi v^2} \right)^2 \sum_{k,l = e,\mu,\tau}  |\mathcal{K}^{U_{ij kl}}_L |^2 +|\mathcal{K}^{D_{ij kl}}_R |^2 + \mathcal{O}(\lambda) .
  \end{aligned}
\end{equation}

\subsection{Dipole operators}
\label{app:WET_dipole}

The effective Hamiltonian for dipole operator is given by 
\begin{align}
  \label{eq:hamiltoniandipole}
  \mathcal{H}^{\text{Dip}}_{\text{eff}}&=-\frac{4\, G_F }{\sqrt{2}}     
    \left(  C_7^{(\prime)}  \mathcal{O}_7^{(\prime)}  +  C_8^{(\prime)}  \mathcal{O}_8^{(\prime)}   \right) \, ,
\end{align}
where the electromagnetic and chromomagnetic operators for $q \to q^\prime$ transitions are given as
\begin{align}
    \mathcal{O}_7^{(\prime)} &=\frac{e}{16\pi^2}m_q \lambda_{CKM} \left(\bar{q}^\prime_{L(R)}\sigma^{\mu\nu} F_{\mu \nu} q_{R(L)}\right)~,\label{eq:O7}\\
    \mathcal{O}_8^{(\prime)} &=\frac{g_s}{16\pi^2}m_q \lambda_{CKM} \left(\bar{q}^\prime_{L(R)} \sigma^{\mu\nu} T^A G^{A}_{\mu \nu} q_{R(L)}\right)~,\label{eq:O8} 
\end{align} 
where $m_q$ denotes the mass of the parent quark, $\sigma^{\mu\nu} = \frac{i}{2} [ \gamma^{\mu},\gamma^{\nu}]$ and $e,g_s$ the electric and QCD coupling constants, respectively.
The tree level matching is given by 
\begin{align}
  C_8^{(\prime)} &= \frac{8 \pi^2 v^2}{\Lambda^2  y_q g_s\lambda_{CKM}}   C_{u G \, , ij(ji) }  \, ,  \\
  C_7^{(\prime)} &= \frac{8 \pi^2 v^2}{\Lambda^2  y_q g_s\lambda_{CKM}} \left(\cos \theta_W   C_{uB \, , ij(ji) }  + \sin \theta_W  C_{uW \, , ij(ji) } \right) \, , 
\end{align}
where $y_q = m_q\sqrt{2} /  v$, $\theta_W $ is the weak mixing angle and $\lambda_{CKM} = V_{tq'}^* V_{q}$ is a CKM factor, that reads $\lambda_{CKM} =V_{tq^\prime}^* V_{tq}$ for  down sector FCNCs $q \to q^\prime$, and
$\lambda_{CKM} =1$  for $c \to u$  for easier comparison with the literature.
Matching the  chirality-flipped WC $C_8^\prime$ corresponds to flipping the quark flavor indices $i,j$ in the SMEFT WC.

\section{Perturbative calculation}
\label{app:pertub}
The perturbative calculation is done for the LO contribution to the $P_T$-spectrum, which includes the processes $q_i(p_1) \bar q_j(p_2) \rightarrow \nu_k(k_1) \bar \nu(k_2) g(k_3)$, $q_i(p_1) g(p_2) \rightarrow \nu_k(k_1) \bar \nu_l(k_2) q_j(k_3)$ and $\bar q_i(p_1) g(p_2) \rightarrow \nu_k(k_1) \bar \nu_l(k_2) \bar q_j(k_3)$.
The last two processes are related through charge conjugation and the former two through crossing.
The partonic mandelstam variables are defined as 
\begin{equation}
  \begin{aligned}
  \hat s &= (p_1 + p_2)^2  \\
  \hat t &= (p_1 - k_3)^2 \\
  \hat u &= (p_2 - k_3)^2 .
\end{aligned}
\end{equation}
The collider variables are defined by 
\begin{equation}
  \begin{aligned}
  P_T^2 = \frac{\hat t \hat u}{\hat s} \\
  \cosh \eta =-\left(\frac{\hat t + \hat u}{2 \sqrt{\hat t \hat u}}  \right) \\
\end{aligned}
\end{equation}
and $q^2$ is the invariant mass of the $\nu \bar \nu$-system $q^{\mu} = k_1^{\mu} + k_2^{\mu}$.
We introduce a parametrization based on Ref.~\cite{Lam:1978pu} and generalize it to include SMEFT effects, which manifestly separates the partonic kinematics, internal propagators and lepton kinematics.
The partonic cross section can be written as  
  \begin{equation}
    \label{eqn:part_xsec}
    \mathrm{d} \hat \sigma = \frac{ \mathcal{A}_C C_F C_A}{8(2 \pi)^5 } \frac{q^2}{\hat s}\sum_{\mathcal{I}_1 \mathcal{I}_2  } \hat H_{\mu \nu}^{\mathcal{I}_1 \mathcal{I}_2 } D^{\mathcal{I}_1 \mathcal{I}_2 } I^{\mu \nu} \mathrm{d}^4q
  \end{equation}
  where $C_F = 4/3, C_A = 3$ and $\mathcal{A}_C$ is a color averaging factor, which reads $1/9$ for $q\bar q$ inital states and $1/24$ for $q g,\bar q g$ initial states.
Furthermore, we define the multi index $\mathcal{I} = \{I,J,Y,Z\}$, which matches operator $\mathcal{I} \in \{ \text{SM},\text{4F},\text{ZP},\text{EW},\text{G} \}$ with explicit vertices $I \in \{ V,D\}$, for vector or dipole copulings, with chirality $Y \pm 1 $ for the four-fermion or $q\bar qZ$-vertex and vertex $J \in \{ V,D\}$ with chirality $Z = \pm 1$ for the $q\bar q g$-vertex.
Formula \eqref{eqn:part_xsec} can be used to explicitly calculate all three partonic processes, however the focus is on the process 
$q_i(p_1) \bar q_j(p_2) \rightarrow \nu_k(k_1) \bar \nu_l(k_2) g(k_3)$ in the following, to derive equation \eqref{eqn:part_xsec}.
The matrix element for the diagram can be written as 
\begin{equation}
  \mathcal{M}_{\mathcal{I}} =  T^A J^{\mu}_{\mathcal{I}}C_{\mathcal{I}} J_{\mu}^{\text{lept}} ,
\end{equation}
where the partonic current is defined as 
  \begin{equation}
    \label{eqn:currents}
     J^{\mu}_{\mathcal{I}}=  i\bar v(p_2)\left( \Gamma^{\mu}_{ I Y }(q) \left(\frac{ \slashed{p_1} - \slashed{k_3}}{\hat t} \right) \Gamma^{\nu}_{ J Z}(k_3) - \Gamma^{\nu}_{ J Z}(k_3) \left(\frac{\slashed{p_2} - \slashed{k_3}}{\hat u}\right) \Gamma^{\mu }_{I Z}(q) \right) u(p_1) \epsilon^{*}_{\nu}(k_3) 
  \end{equation} 
  with vertices 
  \begin{equation}
    \Gamma^{\mu}_{ I Y }(p) =\begin{cases}
      i \gamma^{\mu} \left( \frac{1 + Y \gamma^5}{2} \right)& I = \text{V} \\
      \sigma^{\mu \alpha} p_{\alpha}\left( \frac{1 + Y \gamma^5}{2} \right) & I = \text{D} 
    \end{cases} 
  \end{equation}
The leptonic current is defined as 
\begin{equation}
  J_{\mu}^{\text{lept}} = i \bar u(k_1) \gamma^{\mu} \frac{1 - \gamma^5}{2} v(k_2),
\end{equation}
which is left chiral, since only left handed neutrinos are considered in the final state. 
The coefficient function $\mathcal{C}_{\mathcal{I}}$ includes SM couplings, propagators and WCs, depending on the type of operator insertion $\mathcal{I}$. They are given in Sec.~ \ref{app:coeff_functions}.
The squared Matrix element (summed over all operators, final states and averaged over initial states) reads 
\begin{equation}
\begin{aligned}
   |\overline{\mathcal{M}}|^2 &= \frac{\mathcal{A}_C}{4} \sum_{\mathcal{I}_1 \mathcal{I}_2, \text{spins}}\mathcal{M}_{\mathcal{I}_1} \mathcal{M}_{\mathcal{I}_2}^* \\
  &=\frac{\mathcal{A}_C}{4} \text{Tr} \{ T^A T^A \} \sum_{\mathcal{I}_1 \mathcal{I}_2,\text{spins}} D^{\mathcal{I}_1\mathcal{I}_2} J^{\mu}_{\mathcal{I}_1} (J^{\nu}_{\mathcal{I}_2})^\dagger J_{\mu}^{\text{lept}}(J_{\nu}^{\text{lept}})^\dagger
\end{aligned}
\end{equation}
where $\text{Tr} \{ T^A T^A \} = C_F C_A$ and 
\begin{equation}
    \label{eqn:Dmat}
  D^{\mathcal{I}_1\mathcal{I}_2} = \mathcal{C}_{\mathcal{I}_1} \mathcal{C}_{\mathcal{I}_2}^*, 
\end{equation}
where $\mathcal{C}_{\mathcal{I}_1}$ can be read of the feynman rules for the SMEFT \cite{SMEFT_feyn} and are given in App.~ \ref{app:coeff_functions}.
The three-body phase space can be separated into 
\begin{equation}
  \mathrm{d}\Phi_3 =\frac{1}{(2\pi)^9} \left( \delta^4\left(q -k_2 -k_3\right) \frac{\mathrm{d}^3 k_1}{2 E_1}\frac{\mathrm{d}^3 k_2}{2 E_2} \right) \left( \delta^4\left(p_1+p_2 -q -k_3\right) \frac{\mathrm{d}^3 k_3}{2 E_3} \right) \mathrm{d}^4 q
\end{equation}
which leads to the total partonic cross section
\begin{align}
  \mathrm{d} \hat \sigma =\frac{ \mathcal{A}_C C_F C_A}{8 (2 \pi)^5 }  \frac{q^2}{\hat s}\sum_{\mathcal{I}_1 \mathcal{I}_2  } \hat H_{\mu \nu}^{\mathcal{I}_1 \mathcal{I}_2 } D^{\mathcal{I}_1 \mathcal{I}_2 } I^{\mu \nu} \mathrm{d}^4q.
\end{align}
The parton tensor is defined as  
  \begin{equation}
    \label{eqn:part_tensor}
    \hat H^{\mathcal{I}_1 \mathcal{I}_2 }_{\mu \nu} = \sum_{\text{spins}}  \int \frac{\mathrm{d}^3 k_3}{ 2 E_3} \delta^4\left( p_1 + p_2 - q - k_3\right)   J^{\mu}_{\mathcal{I}_1} \left(J^{\nu}_{\mathcal{I}_2}\right)^{\dagger}  
\end{equation}
while the lepton tensor can be calculated
\begin{equation}
  \begin{aligned}
  \label{eqn:lepton_tensor}
  I_{\mu \nu} = &\int \frac{\mathrm{d}^3 k_1}{2 E_1}\frac{\mathrm{d}^3 k_2}{2 E_2} \frac{1}{q^2}\sum_{\text{spins}}   J_{\mu}^{\text{lept}}  (J_{\nu}^{\text{lept}})^\dagger \\
  =& \frac{\pi}{3} \left( - \eta^{\mu \nu} + \frac{q^{\mu} q^{\nu} }{q^2} \right).
\end{aligned}
\end{equation}
The introduced separation in equation \ref{eqn:part_xsec} is useful, since all relevant insertion of SMEFT operators are included fully in the partonic part $\hat H^{\mathcal{I}_1 \mathcal{I}_2 }_{\mu \nu} $, which in the end has to be contracted with the lepton tensor to produce the final cross section.
Therefore, it is useful to further analyse the properties, which have been worked out in reference \cite{Lam:1978pu} for the hadronic counter part.
This leads to the parametrization 
\begin{equation}
  \label{fig:parton_param}
  \begin{aligned}
  \hat H^{\mathcal{I}_1 \mathcal{I}_2 }_{\mu \nu} =&\delta(\hat s + \hat t +\hat u -q^2)\frac{E_q}{ E_3}  \bigg\{ \eta^{\perp}_{ \mu \nu} \hat F^{\mathcal{I}_1 \mathcal{I}_2 }_1(\hat s,\hat t,\hat u) + \left(p^\perp_{1}+ p^\perp_{2} \right)_{\mu  } \left(p^\perp_{1}+ p^\perp_{2} \right)_{\nu  }  \hat F^{\mathcal{I}_1 \mathcal{I}_2}_2(\hat s,\hat t,\hat u) \\
  &+  p^{\perp}_{1  \mu  } p^{\perp}_{1  \nu}  \hat F^{\mathcal{I}_1 \mathcal{I}_2}_3(\hat s,\hat t,\hat u)  + p^\perp_{2 \mu  } p^\perp_{2  \nu}  \hat F^{\mathcal{I}_1 \mathcal{I}_2}_4(\hat s,\hat t,\hat u)\\
  &+ \left(\eta_{\mu \nu} -\frac{2}{\hat s}\left(p_{1 \mu} p_{2 \nu} + p_{1\nu} p_{2\mu} \right)\right) \hat F^{\mathcal{I}_1 \mathcal{I}_2}_5(\hat s,\hat t,\hat u) +  q_{\mu}\text{ -terms} \bigg\},
\end{aligned}
\end{equation}
where the energy momentum conserving $\delta$-function and a phase space factor have been factored.
The lorentz structures are given by 
\begin{align*}
  \eta^{\perp}_{ \mu \nu} = \eta_{\mu \nu } - \frac{q_{\mu} q_{\nu}}{q^2} \\
  p^{\perp}_{i,\mu} = \eta_{\mu \nu } \frac{p^{\nu}_i}{\sqrt{\hat s}},
\end{align*}
following Ref.~\cite{Lam:1978pu}.
We assume $ \hat H^{\mathcal{I}_1 \mathcal{I}_2 }_{\mu \nu}$ to be symmetric and ignore terms proportional to $q^{\mu}$, since $I^{\mu \nu}$ is manifestly symmetric, as well as $I^{\mu \nu}q_{\mu} =0 $.
The structure functions $F^{\mathcal{I}_i \mathcal{I}_2 }_1(\hat s,\hat t,\hat u)$ capture all relevant contributions as in reference \cite{Lam:1978pu}.
However, we need to include one additional piece, since the general current $J^{\mu}_{\mathcal{I}}$ is not conserved and therefore $\hat H^{\mathcal{I}_1 \mathcal{I}_2 }_{\mu \nu} q^{\mu} \neq 0$, which leads to the inclusion of $F^{\mathcal{I}_1 \mathcal{I}_2}_5(\hat s,\hat t,\hat u)$.
The end result, which directly contributes to the total cross section is given by the contraction
\begin{align}
  \hat H^{ \mathcal{I}_1 \mathcal{I}_2}_{\mu \nu} I^{\mu \nu} &=  \delta(q^2 - \hat s - \hat t - \hat u   ) \frac{E_q}{ E_3} \mathcal{F}^{\mathcal{I}_1\mathcal{I}_2}(\hat s , \hat t, \hat u) ,
\end{align}
where 
\begin{equation}
  \begin{aligned}
  \mathcal{F}^{\mathcal{I}_1\mathcal{I}_2}(\hat s , \hat t, \hat u) &=  \frac{\pi}{3}\bigg( -3 F_1^{\mathcal{I}_1 \mathcal{I}_2 }(\hat s, \hat t ,\hat u) +\frac{1}{4 \hat s (  \hat s + \hat t + \hat u )}\big \{ (\hat t + \hat u)^2 F_2^{\mathcal{I}_1 \mathcal{I}_2}(\hat s, \hat t ,\hat u) \\
&+ (\hat t + \hat s)^2 F_3^{\mathcal{I}_1 \mathcal{I}_2}(\hat s, \hat t ,\hat u) \\
&+  (\hat s + \hat u)^2 F_4^{\mathcal{I}_1 \mathcal{I}_2 }(\hat s, \hat t ,\hat u)+  (2 \hat s ( \hat s + \hat t + \hat u) - \hat u \hat t ) F_5^{\mathcal{I}_1 \mathcal{I}_2}(\hat s, \hat t ,\hat u) \big\} \bigg) .
\end{aligned}
\end{equation} 
The factor 
\begin{equation}
  \frac{E_q}{ E_3} = \frac{\sqrt{q^2 + P_T^2 \cosh^2\eta }}{ P_T \cosh \eta}
\end{equation}
is a flux factor from the $d^3k_3$ integration, which will cancel in the end.
Switching to collider variables ($ \mathrm{d}^4q \rightarrow \pi\frac{E_3}{E_q} P_T \mathrm{d} P_T \mathrm{d}q^2 \mathrm{d}\eta   $), summing over neutrino flavors, integrating over $\eta$  with the $\delta$-function, the hard cross sections reads
\begin{equation}
\begin{aligned}
  \label{eqn:parton_xsec}
  \frac{\mathrm{d}\hat \sigma( q_i \bar q_j \rightarrow \nu \bar \nu g)}{\mathrm{d} P_T \mathrm{d} q^2}  &=\frac{\mathcal{A}_C C_F C_A}{16  (2\pi)^4} \frac{P_T q^2 }{\hat s }\sum_{k,l}    \sum_{\mathcal{I}_1\mathcal{I}_2} \int \mathrm{d} \eta \delta\left( q^2 - \hat s +2 P_T \sqrt{\hat s} \cosh\eta \right) D^{\mathcal{I}_1\mathcal{I}_2}(q^2)\mathcal{F}^{\mathcal{I}_1\mathcal{I}_2}(\hat s , \hat t, \hat u)\\
  &= \frac{\mathcal{A}_C C_F C_A}{16  (2\pi)^4}\frac{P_T q^2}{\hat s }\sum_{k,l} \sum_{\mathcal{I}_1\mathcal{I}_2} \frac{\Theta \left( \hat s -q^2 + 2 P_T \sqrt{\hat s}\right)}{\sqrt{ (\hat s-q^2)^2 -4 P_T^2 \hat s}} D^{\mathcal{I}_1\mathcal{I}_2}(q^2) \tilde{\mathcal{F}}^{\mathcal{I}_1\mathcal{I}_2}(\hat s , \hat t, \hat u),
\end{aligned}
\end{equation}
where 
\begin{equation}
  \tilde{\mathcal{F}}^{\mathcal{I}_1\mathcal{I}_2}(\hat s , \hat t, \hat u) =   \left(\mathcal{F}^{\mathcal{I}_1\mathcal{I}_2}(\hat s , \hat t, \hat u) +\mathcal{F}^{\mathcal{I}_1\mathcal{I}_2}(\hat s , \hat u, \hat t) \right)
\end{equation}
is the symmetrized over $\hat t, \hat u$, since the antisymmetric part vanishes upon performing the $\eta$-integration.
The mandelstam variables now are given as a function of $P_T, q^2,\hat s$ and read 
\begin{align}
  \hat t &= \frac{1}{2} \left( \sqrt{(q^2 -\hat s )^2 -4 P_T^2 \hat s} + q^2 -\hat s \right) \\
  \hat u &= \frac{1}{2} \left( -\sqrt{(q^2 -\hat s )^2 -4 P_T^2 \hat s} + q^2 -\hat s \right).
\end{align}
The $P_T$ spectrum can then be obtained by performing the $q^2 $ integration.
The $qg$ cross sections can be calculated by crossing $\hat s \Leftrightarrow \hat t$ and an overall minus sign in the function $\mathcal{F}^{\mathcal{I}_1\mathcal{I}_2}$.
\subsection{ Structure functions}
In the following section all structure functions as defined in equation \eqref{fig:parton_param} are displayed. 
These are sorted into four categories: The purely vector couplings ($I_1 = I_2 = J_1 = J_2 = V$), the gluon dipole couplings ($I_1 = I_2 = \text{V}, J_1 = J_2 = \text{D}$), the EW dipole corrections ($I_1 = I_2 = \text{D}, J_1 = J_2 = \text{V})$.
Note that there are also additional combinations, which capture the interference between gluon and EW dipole corrections, which are not listed, since there are not relevant for this work.  
\subsubsection{Vector}
The structure functions read:
\begin{align}
  F_1^{\text{V}\text{V}Y_1Z_1\text{V}\text{V}Y_2Z_2 } &= \frac{ -2 \left( (\hat s + \hat t)^2 + ( \hat s + \hat u)^2 \right) }{ \hat t \hat u} \delta_{Y_1,Y_2}\delta_{Y_1,Z_1} \delta_{Y_2,Z_2}\\
  F_2^{\text{V}\text{V}Y_1Z_1\text{V}\text{V}Y_2Z_2 } &= F_5^{\text{V}\text{V}Y_1Z_1\text{V}\text{V}Y_2Z_2  } = 0 \\
  F_3^{\text{V}\text{V}Y_1Z_1\text{V}\text{V}Y_2Z_2  } &= F_4^{\text{V}\text{V}Y_1Z_1\text{V}\text{V}Y_2Z_2 } = \frac{-8 \hat s  \left( \hat s + \hat t + \hat u \right) }{\hat t \hat u}\delta_{Y_1,Y_2}\delta_{Y_1,Z_1} \delta_{Y_2,Z_2}
\end{align}
\subsubsection{Gluon Dipole}
The structure functions read:
\begin{align}
  F_1^{\text{V} \text{D}  Y_1 Z_1 \text{V}  \text{D} Y_2 Z_2} &= -4 \hat s \delta_{Y_1 ,- Y_2} \delta_{Z_1 ,Z_2}\\
  F_2^{\text{V} \text{D}  Y_1 Z_1 \text{V}  \text{D} Y_2 Z_2} &= -4 \frac{\hat s}{\hat t \hat u} \left( \hat s^2 + \hat s ( \hat t + \hat u) - \hat t \hat u\right)  \delta_{Y_1 ,- Y_2} \delta_{Z_1 ,Z_2} \\
  F_3^{\text{V} \text{D}  Y_1 Z_1 \text{V}  \text{D} Y_2 Z_2}  &= -4 \frac{\hat s}{\hat t \hat u} \left( \hat u^2+ \hat u (\hat t +\hat s ) - \hat s \hat t \right)  \delta_{Y_1 ,- Y_2} \delta_{Z_1 ,Z_2} \\
  F_4^{\text{V} \text{D}  Y_1 Z_1 \text{V}  \text{D} Y_2 Z_2}  &= -4 \frac{\hat s}{\hat t \hat u}\left( \hat t^2+ \hat t (\hat u  +\hat s ) - \hat s \hat u \right)  \delta_{Y_1 ,- Y_2} \delta_{Z_1 ,Z_2} \\
  F_5^{\text{V} \text{D}  Y_1 Z_1 \text{V}  \text{D} Y_2 Z_2} &= - 2 \hat s  \delta_{Z_1,Z_2}  Y_1 Y_2 Z_1 Z_2 
\end{align}
\subsubsection{EW Dipole}
The structure functions read:
\begin{align}
  F_1^{\text{D} \text{V} Y_1 Z_1 \text{D} \text{V}  Y_2 Z_2} &= -2 \hat s \delta_{Y_1 , Y_2} \delta_{Z_1 ,Z_2}\\
  F_2^{\text{D} \text{V} Y_1 Z_1 \text{D} \text{V}  Y_2 Z_2} &= 4 \hat s \frac{\left(\hat s + \hat t +\hat u\right)}{\hat t \hat u} \delta_{Y_1,Y_2}\left( \hat u  \delta_{Z_1,Z_2} \delta_{Y_1,Z_1} + \hat t  \delta_{Z_1,Z_2} \delta_{Y_1,-Z_1}   -  \left( \hat s + \hat t + \hat u\right) \delta_{Z_1,-Z_2} \right)  \\
  F_3^{\text{D} \text{V} Y_1 Z_1 \text{D} \text{V}  Y_2 Z_2}  &= 4 \hat s \frac{\left(\hat s + \hat t +\hat u\right)}{\hat t \hat u} \delta_{Y_1,Y_2}\left( -\hat u  \delta_{Z_1,Z_2} \delta_{Y_1,Z_1} + \hat t  \delta_{Z_1,Z_2} \delta_{Y_1,-Z_1}   +  \left( \hat s + \hat t + \hat u\right) \delta_{Z_1,-Z_2} \right)  \\
  F_4^{\text{D} \text{V} Y_1 Z_1 \text{D} \text{V}  Y_2 Z_2}  &= 4 \hat s \frac{\left(\hat s + \hat t +\hat u\right)}{\hat t \hat u} \delta_{Y_1,Y_2}\left( \hat u  \delta_{Z_1,Z_2} \delta_{Y_1,Z_1} - \hat t  \delta_{Z_1,Z_2} \delta_{Y_1,-Z_1}   +  \left( \hat s + \hat t + \hat u\right) \delta_{Z_1,-Z_2} \right)  \\
  F_5^{\text{D} \text{V} Y_1 Z_1 \text{D} \text{V}  Y_2 Z_2}  &= 0 
\end{align}

\subsection{Coefficient functions}
\label{app:coeff_functions}
The $D^{\mathcal{I}_1\mathcal{I}_2}(q^2)$ function can be constructed from the feynman rules in reference \cite{SMEFT_feyn}.
They factorizes into two parts as can be seen in equation \eqref{eqn:Dmat} and depends on $C_{\mathcal{I}}$, which are listed in the following. 
The results are given in terms of the $SU(2)_L$, $U(1)_Y$ couplings $g_1$, $g_2$, respectively, as well as the $SU(3)_c$ coupling $g_s$,  the $Z$ mass $M_Z$ and its width $\Gamma$.
Note that all expressions are given in the flavor basis.
The SM $Z q \bar q$ couplings read 
\begin{align}
    \epsilon_L^{i,j} &= \begin{cases}
        \frac{ g_2^2 -3 g_1^2}{6\sqrt{g_1^2 + g_2^2}} & i,j \text{ Up-sector} \\
        \frac{ g_2^2 +3 g_1^2}{6\sqrt{g_1^2 + g_2^2}} & i,j \text{ Down-sector}
    \end{cases}
     \\
     \epsilon_R^{i,j} &= \begin{cases}
        \frac{ 4 g_2^2}{6\sqrt{g_1^2 + g_2^2}}                          & i,j \text{ Up-sector} \\
        \frac{ -2 g_2^2}{6\sqrt{g_1^2 + g_2^2}}   & i,j \text{ Down-sector}.
     \end{cases} 
\end{align}
The couplings $\mathcal{C}^{\mathcal{I}}$ implicitly depend on the flavors $i,j,k,l$ and on the chiralities $Y,Z$ of the $Zq\bar q$ ($q\bar q \nu \bar \nu$ for 4F)-vertex or $g q \bar q$-vertex, respectively.
Furthermore, difference appear through the different sectors, i.e. up- or down-sector. This lead to some signs, where the upper(lower) one corresponds to the up(down)-sector. 
\begin{align}
    \mathcal{C}^{SM}  &= g_s  \delta_{i,j}\delta_{k,l} \frac{\sqrt{g_1^2 + g_2^2}}{2} \frac{1}{q^2 - M_Z^2 + i \Gamma M_Z} \times
             \begin{cases}
                \epsilon_L^{i,j} &  Y = +1, Z = \pm 1 \\
                \epsilon_R^{i,j} &  Y = -1, Z = \pm 1
            \end{cases} \\
    \mathcal{C}^{4F}  &= g_s \times \begin{cases}
        C^{lu(d)}_{klij} & Y = +1,Z = \pm 1 \\
       \left(C_{lq,klij}^1 \pm C_{lq,klij}^{3}\right) &Y = -1,Z = \pm 1 \\
    \end{cases}\\
    \mathcal{C}^{ZP}  &=  \frac{i g_s v^2 \sqrt{g_1^2 +g_2^2}}{2} \frac{ \delta_{k,l}}{q^2 - M_Z^2 + i \Gamma M_Z} 
    \times \begin{cases}
            C_{\phi u(d)}  & Y = +1,Z = \pm 1 \\
            \left(C_{\phi q , m n }^1 \mp C_{\phi q , m n }^3\right) &  Y = -1,Z = \pm 1
    \end{cases} \\
    \mathcal{C}^{EW} &= - \frac{\sqrt{2}  v}{\sqrt{g_1^2 + g_2^2}}\frac{1}{q^2 - M_Z^2 + i \Gamma M_Z} \times
    \begin{cases}
          g_1 C^{u(d)W}_{i,j} -  g_2 C^{u(d)B}_{i,j}  & Y = +1,Z = \pm 1 \\
          g_1 C^{u(d)W}_{j,i} -  g_2 C^{u(d)B}_{j,i}  & Y = -1,Z = \pm 1
    \end{cases} \\ 
    \mathcal{C}^{G} &= -  \sqrt{2} v \frac{\sqrt{g_1^2 +g_2^2}}{2} \delta_{k,l} \frac{1}{q^2 - M_Z^2 + i \Gamma M_Z}  \times 
    \begin{cases}
        C^{u(d)G}_{j,i}\epsilon_R^{i,j}  &   Y = +1, Z = +1 \\
        C^{u(d)G}_{j,i}\epsilon_L^{i,j}  &   Y = -1, Z = +1 \\
        {C^{u(d)G}_{i,j}}^*\epsilon_R^{i,j}  &   Y = +1, Z = -1 \\
        {C^{u(d)G}_{i,j}}^*\epsilon_L^{i,j}  &   Y = -1, Z = -1 \\
    \end{cases}
\end{align}

\subsection{Cross sections}
\begin{figure}
  \centering 
  \begin{subfigure}{0.48 \textwidth}
    \includegraphics[width = \linewidth]{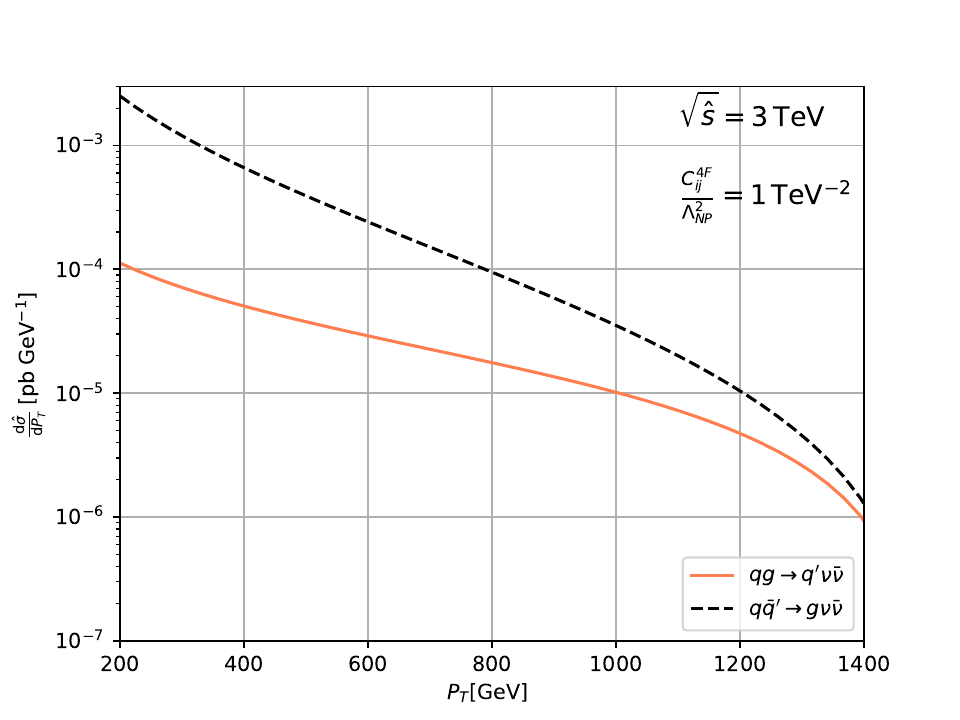}
    \subcaption{}
  \end{subfigure}
  \begin{subfigure}{0.48 \textwidth}
    \includegraphics[width = \linewidth]{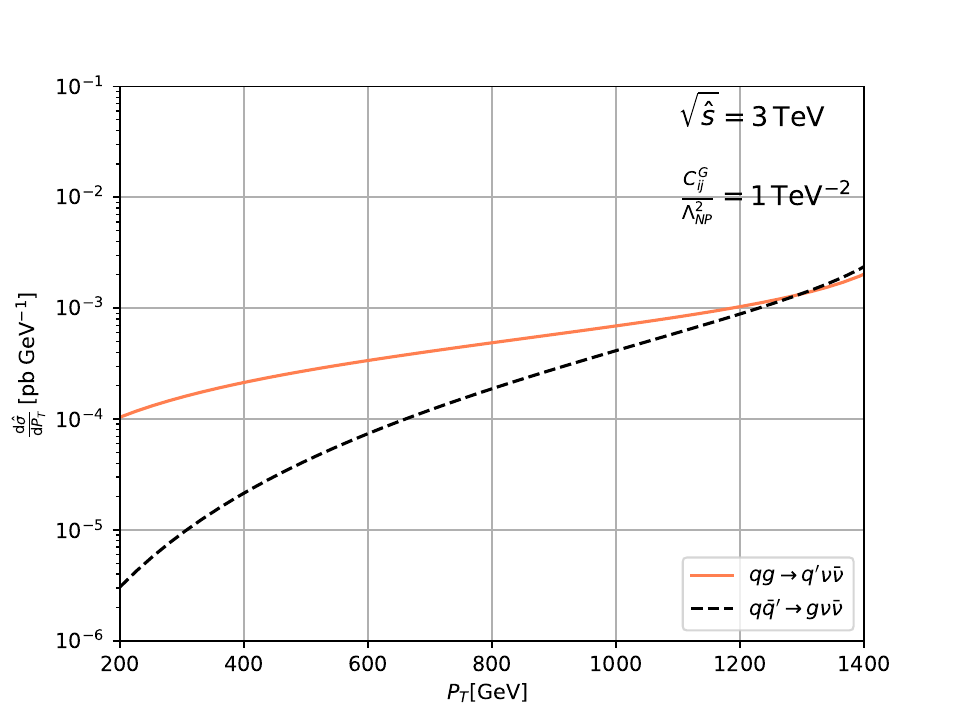}
    \subcaption{}
  \end{subfigure}
  \begin{subfigure}{0.48 \textwidth}
    \includegraphics[width = \linewidth]{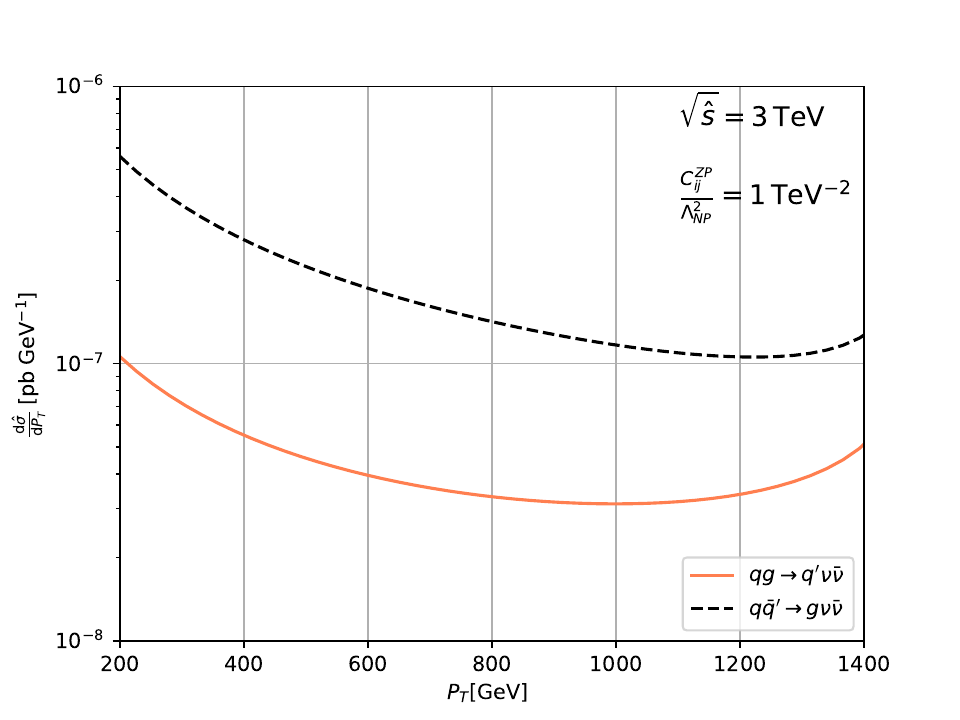}
    \subcaption{}
  \end{subfigure}
  \begin{subfigure}{0.48 \textwidth}
    \includegraphics[width = \linewidth]{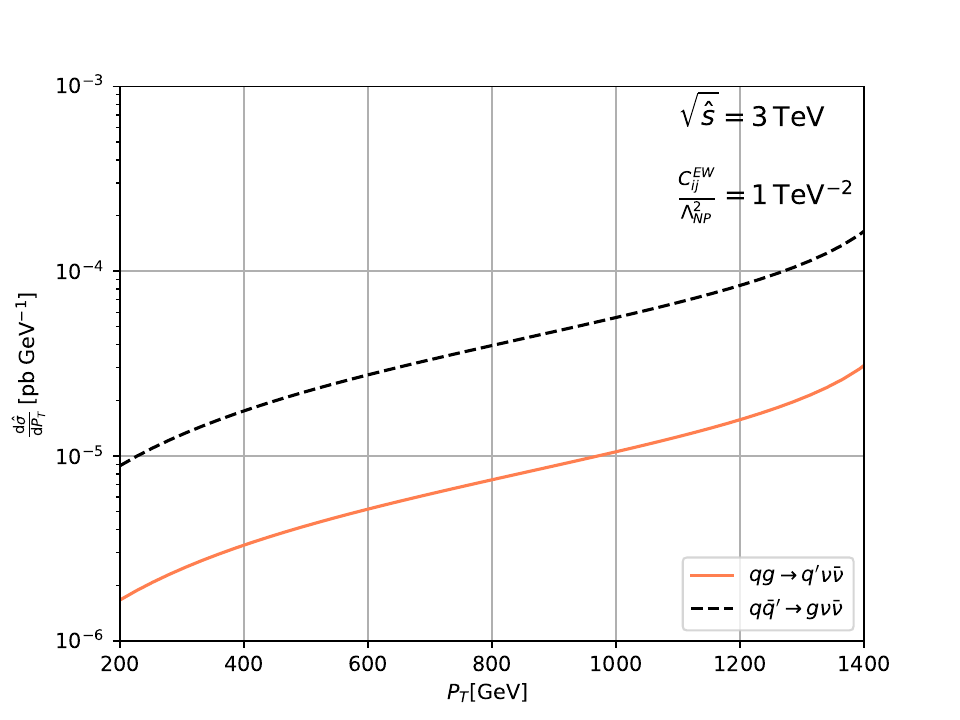}
    \subcaption{}
  \end{subfigure}
  \caption{Comparison of partonic cross sections of  $qg$ (orange, solid) and $q\bar q$ (black, dashed) contributions 
   for the effective WCs $C^{4F}_{ij} $ (a), $C^{G}_{ij} $ (b), $C^{ZP}_{ij} $ (c) and $C^{EW}_{ij} $ (d) as defined in Secs.~\ref{sec:fourfermion},\ref{sec:gluon-dipole} and \ref{sec:penguin}.
  Note that in (b) the difference between up- and down-sector is not shown, since the difference is below $1 \%$.  }
  \label{fig:parton_xsec}
\end{figure}

In the following subsection, the differential cross sections for the SM and all operator insertions are given. 
These are based on formula \eqref{eqn:parton_xsec} and are integrated over $q^2$.
The $P_T$-spectra for all operators including a $Z$-boson propagator, are given in the Narrow width approximation (NWA). 
Explicitly this reads 
\begin{equation}
    \frac{1}{ (q^2 -M_Z^2)^2 +\Gamma^2 M_Z^2} \rightarrow \delta\left( q^2 -M_Z^2\right) \frac{\pi}{\Gamma M_Z},
\end{equation}
which trivializes the $q^2$ integration in equation \eqref{eqn:parton_xsec} and therefore allows the consideration of explicit analytical results.
The couplings in the previous section are translated into the input parameters $ v, M_Z, \alpha_s , M_W$.
High energy results are obtained by the expansion $M_Z \sim M_W \ll P_T \leq\sqrt{\hat s}/2 $, which can then be written in terms of the scaling variable $ x = 2 P_T/ \sqrt{\hat s}\leq1$.
The SM $Z q \bar q$ couplings now read 
\begin{align}
  \label{eqn:SM_Z_couplings}
    \epsilon_L^{i,j} &= \begin{cases}
         \frac{\left(M_Z^2 - 4 M_W^2\right)}{3 M_Z v } \qquad \text{for}\; i,j \text{ Up-sector} \\
         \frac{\left(M_Z^2 + 2 M_W^2\right)}{3 M_Z v } \qquad \text{for}\; i,j \text{ Down-sector}
    \end{cases}
     \\
     \epsilon_R^{i,j} &= \begin{cases}
         \frac{4\left(M_Z^2 -  M_W^2\right)}{3 M_Z v} \qquad \text{for}\; i,j \text{ Up-sector} \\
         \frac{2\left(-M_Z^2 +  M_W^2\right)}{3 M_Z v} \qquad \text{for}\; i,j \text{ Down-sector}
     \end{cases} 
\end{align}
All NWA results are given in terms of
\begin{equation}
     \text{Br}( Z \rightarrow \nu \bar \nu) = \frac{G_F M_Z^3}{3 \sqrt{2} \pi \Gamma}  ,         
\end{equation}
where   $\text{Br}( Z \rightarrow \nu \bar \nu)\simeq 20 \%$ \cite{ALEPH:2005ab}.
In Fig.~\ref{fig:parton_xsec} all  partonic cross sections are shown to allow for a comparison of
$qg$ and $\bar q q$ contributions at parton level.

\begin{figure}
  \centering 
  \begin{subfigure}{0.48 \textwidth}
    \includegraphics[width = \linewidth]{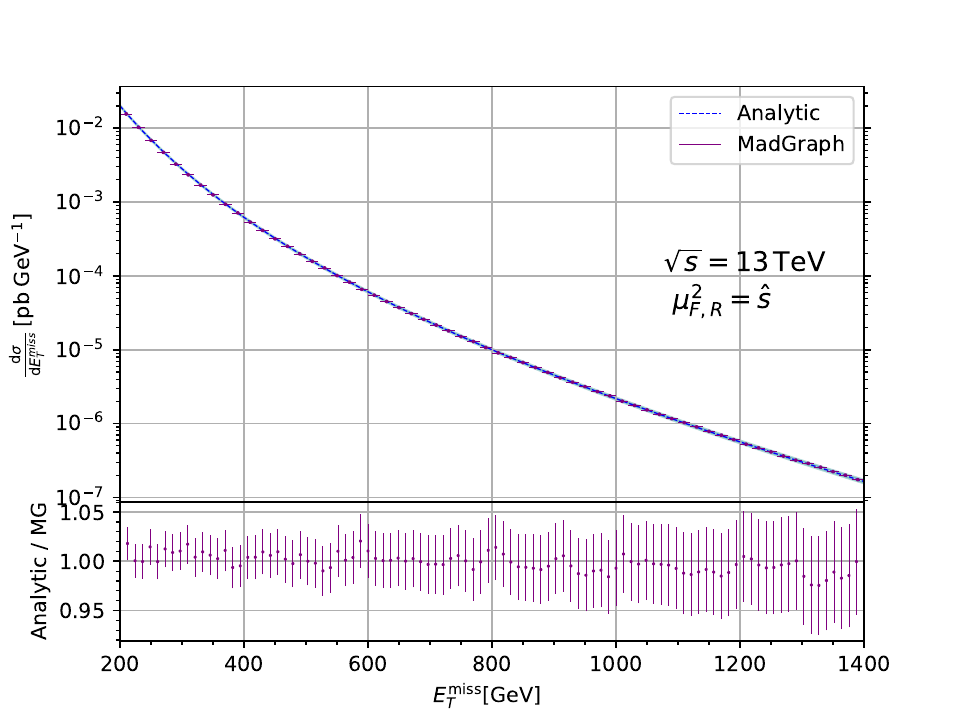}
    \subcaption{}
  \end{subfigure}
  \begin{subfigure}{0.48 \textwidth}
    \includegraphics[width = \linewidth]{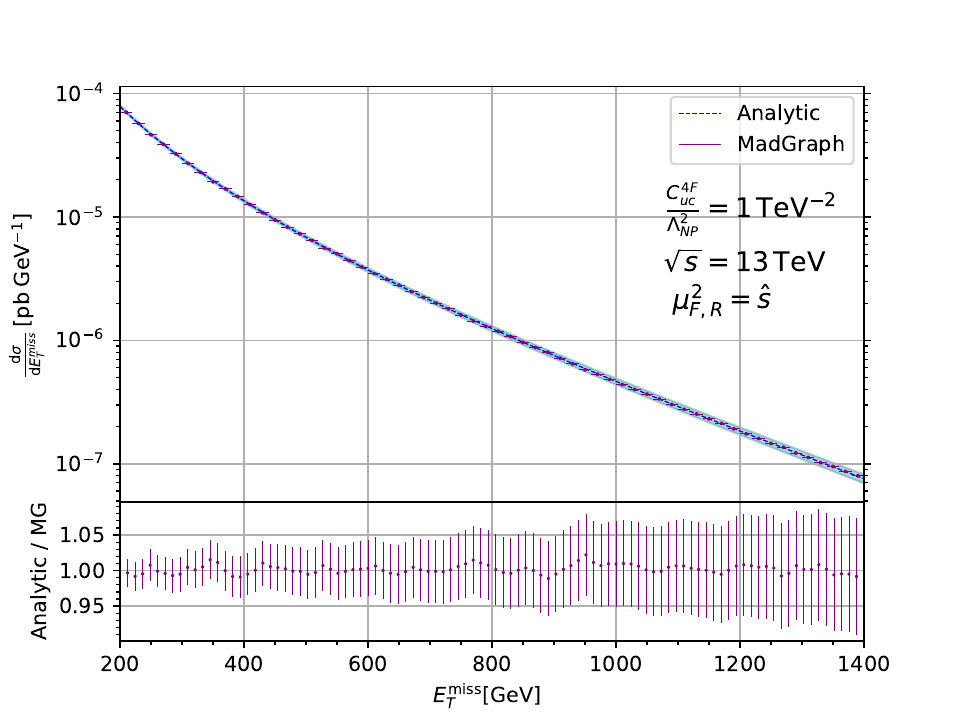}
    \subcaption{}
  \end{subfigure}
  \begin{subfigure}{0.48 \textwidth}
    \includegraphics[width = \linewidth]{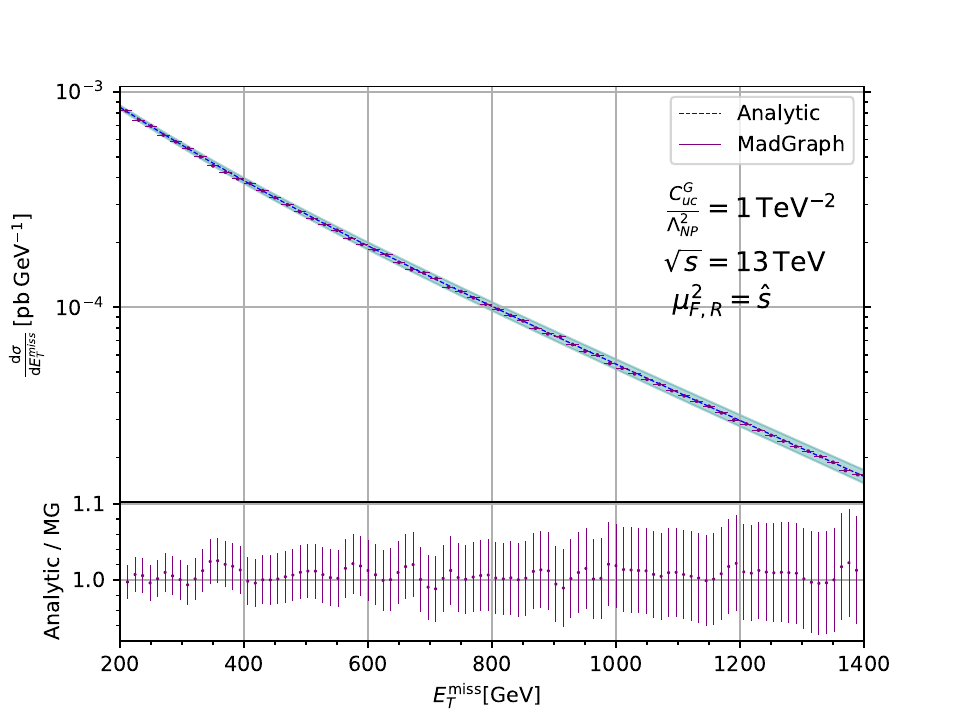}
    \subcaption{}
  \end{subfigure}
  \begin{subfigure}{0.48 \textwidth}
    \includegraphics[width = \linewidth]{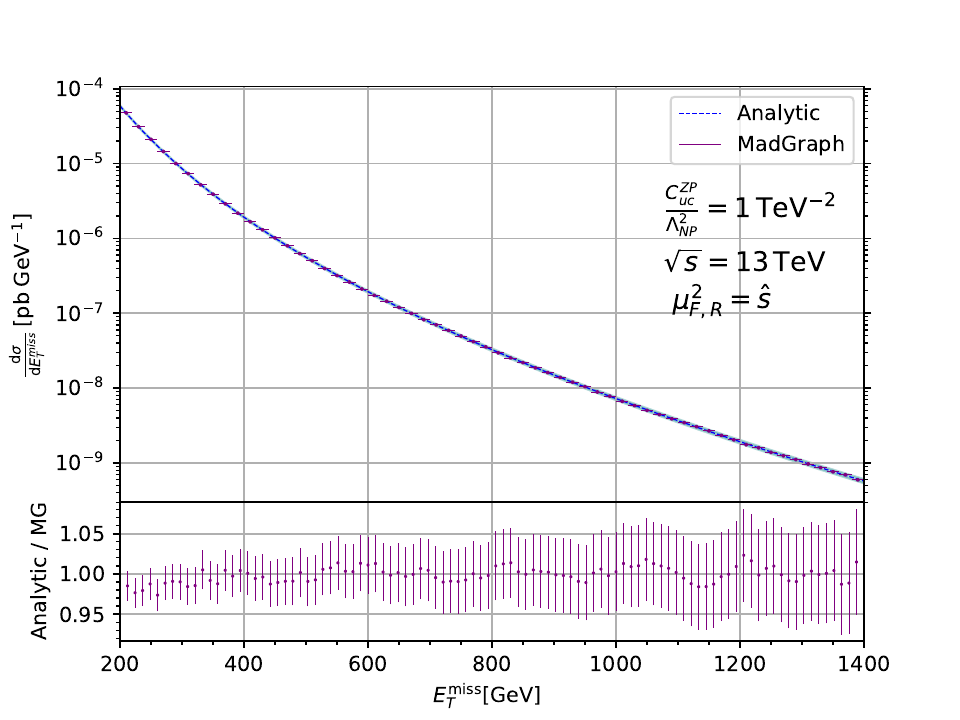}
    \subcaption{}
  \end{subfigure}
  \begin{subfigure}{0.48 \textwidth}
    \includegraphics[width = \linewidth]{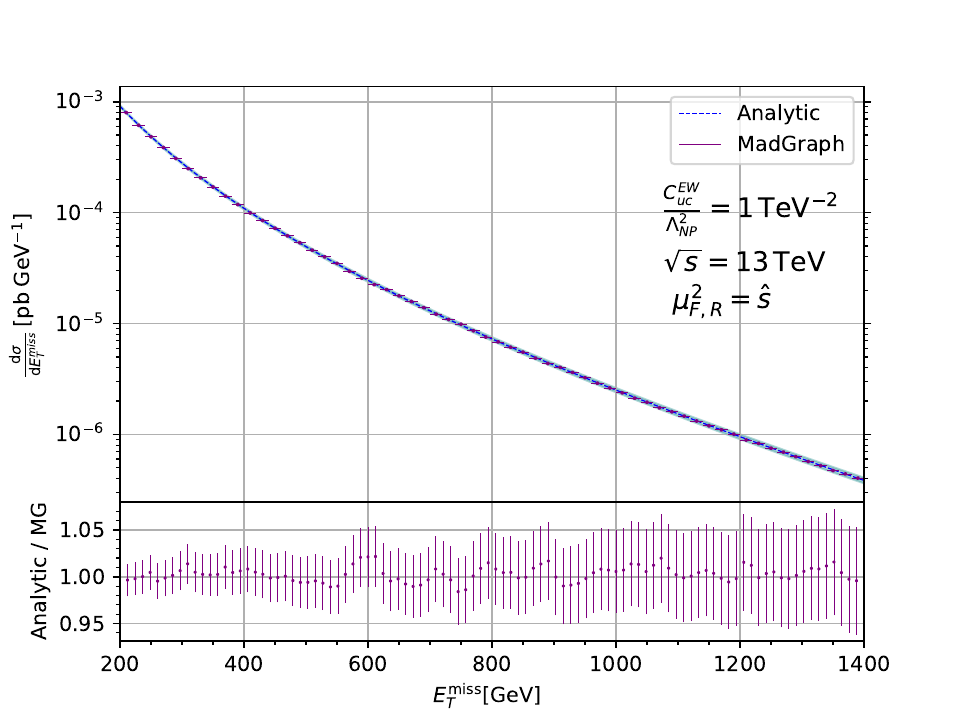}
    \subcaption{}
  \end{subfigure}
  \caption{Analytical results from App.~\ref{app:pertub}, compared with the \texttt{MadGraph5\_aMC@NLO} \cite{MG5} results using the UFO model \texttt{SMEFTsim\_general\_MwScheme\_UFO} \cite{SMEFTsim}.
  Shown are the SM prediction (a), as well as predictions for $C^{4F}_{uc} $ (b), $C^{G}_{uc} $ (c), $C^{ZP}_{uc} $ (d) and $C^{EW}_{uc}$ (e), see Secs.~\ref{sec:fourfermion},\ref{sec:gluon-dipole} and \ref{sec:penguin}. }
  \label{fig:validation}
\end{figure}
\subsubsection{SM prediction}
The $P_T$-spectra are then given in the NWA 
\begin{align}
    \frac{\mathrm{d}\hat \sigma( q_i \bar q_j \rightarrow \nu \bar \nu g)}{\mathrm{d} P_T } &= \frac{ 4 \alpha_s  \left( {\epsilon_L^{i,j}}^2 + {\epsilon_R^{i,j}}^2   \right) \text{Br}\left( Z \rightarrow \nu \bar \nu\right) }{9   }   \frac{ M_Z^4 + \hat s ( \hat s -2 P_T^2) }{ P_T \hat s^2\sqrt{(\hat s -M_Z^2)^2 - 4 P_T^2 \hat s} } \\
    \frac{\mathrm{d}\hat \sigma( q_i g \rightarrow \nu \bar \nu q_j)}{\mathrm{d} P_T } &= -\frac{ \alpha_s   \left( {\epsilon_L^{i,j}}^2 + {\epsilon_R^{i,j}}^2   \right)\text{Br}\left( Z \rightarrow \nu \bar \nu\right) }{12   } \\ \nonumber
    &\times  \frac{ 2 M_Z^6 - 4 M_Z^2 \hat s + 3 M_Z^2 \hat s( \hat s - P_T^2) - \hat s^2 (P_T^2 + \hat s)}{ P_T \hat s^3 \sqrt{(\hat s -M_Z^2)^2 - 4 P_T^2 \hat s} } \\ 
\end{align}
In the high energy limit these approach
\begin{align}
    \frac{\mathrm{d}\hat \sigma( q_i \bar q_j \rightarrow \nu \bar \nu g)}{\mathrm{d} P_T } &\approx  \frac{ 4 \sqrt{2} \alpha_s  \text{Br}( Z \rightarrow \nu \bar \nu) \left( {\epsilon_L^{i,j}}^2 + {\epsilon_R^{i,j}}^2   \right) }{9 } \frac{1}{\hat s^{3/2}}\frac{ 2  -x^2 }{ x  \sqrt{1-x^2} } \\
    \frac{\mathrm{d}\hat \sigma( q_i g \rightarrow \nu \bar \nu q_j)}{\mathrm{d} P_T }      &\approx  \frac{ \alpha_s  \text{Br}( Z \rightarrow \nu \bar \nu) \left( {\epsilon_L^{i,j}}^2 + {\epsilon_R^{i,j}}^2   \right)}{12 \sqrt{2}}\frac{1}{\hat s^{3/2} } \frac{ x^2 + 4 }{ x \sqrt{ 1- x^2} } 
\end{align}
\subsubsection{Four-fermion operators }
The results for 4F operators can be integrated analytically and the $P_T$-spectrum can be given in terms of the scaling variable $ x =  2P_T/\sqrt{\hat s}$:
\begin{align}
    \frac{\mathrm{d}\hat \sigma( q_i \bar q_j \rightarrow \nu \bar \nu g)}{\mathrm{d} P_T } &=  \frac{\alpha_s {C^{4F}_{ij} }^2}{162 \pi^2 \Lambda^4}\sqrt{\hat s} \; \frac{  6 ( x^2 +2 ) \cosh^{-1}\left(x^{-1} \right) - ( x^2 + 17) \sqrt{ 1 - x^2}  }{x} \\
    \frac{\mathrm{d}\hat \sigma( q_i g \rightarrow \nu \bar \nu q_j)}{\mathrm{d} P_T } &= \frac{\alpha_s {C^{4F}_{ij} }^2}{3456 \pi^2 \Lambda^4}\sqrt{\hat s} \; \frac{ 3 x^2 (3 x^2 +4 ) \cosh^{-1}\left(x^{-1} \right) - ( 13 x^2 + 8) \sqrt{ 1 - x^2}  }{x}
\end{align}
For $P_T \sim \sqrt{\hat s}/2$ ($x \rightarrow 1$) these read 
\begin{align}
\frac{\mathrm{d}\hat \sigma( q_i \bar q_j \rightarrow \nu \bar \nu g)}{\mathrm{d} P_T } &\approx   \frac{\sqrt{2}\alpha_s {C^{4F}_{ij} }^2}{81 \pi^2 }\frac{1}{\Lambda^4} \sqrt{\hat s} (1-x)^{3/2} \\
\frac{\mathrm{d}\hat \sigma( q_i g \rightarrow \nu \bar \nu q_j)}{\mathrm{d} P_T }      &\approx   \frac{5\alpha_s {C^{4F}_{ij} }^2}{432 \sqrt{2} \pi^2 } \frac{1}{\Lambda^4}\sqrt{\hat s} (1-x)^{3/2}\;
\end{align}
\subsubsection{Z-penguin operators}
The $P_T$-spectra are then given in the NWA
\begin{align}
    \frac{\mathrm{d}\hat \sigma( q_i \bar q_j \rightarrow \nu \bar \nu g)}{\mathrm{d} P_T } &= \frac{ 4 \alpha_s M_Z^2 v^2  { C^{ZP}_{i,j} }^2 \text{Br}\left( Z \rightarrow \nu \bar \nu\right)}{9 \Lambda^4 } \frac{ M_Z^4 + \hat s ( \hat s -2 P_T^2) }{ P_T \hat s^2\sqrt{(\hat s -M_Z^2)^2 - 4 P_T^2 \hat s}} \\
    \frac{\mathrm{d}\hat \sigma( q_i g \rightarrow \nu \bar \nu q_j)}{\mathrm{d} P_T } &= -\frac{ \alpha_s M_Z^2   { C^{ZP}_{i,j} }^2  \text{Br}\left( Z \rightarrow \nu \bar \nu\right)}{12 \sqrt{2} G_F \Lambda^4} \\ \nonumber
    &\times     \frac{ 2 M_Z^6 - 4 M_Z^2 \hat s + 3 M_Z^2 \hat s( \hat s - P_T^2) - \hat s^2 (P_T^2 + \hat s)}{ P_T \hat s^3 \sqrt{(\hat s -M_Z^2)^2 - 4 P_T^2 \hat s} } 
\end{align}
In the high energy limit these approach
\begin{align}
  \frac{\mathrm{d} \hat \sigma( q_i \bar q_j \rightarrow \nu \bar \nu g)}{\mathrm{d} P_T } &\approx \frac{4  \alpha_s {C^{ZP}_{i,j} }^2 \text{Br}\left( Z \rightarrow \nu \bar \nu\right) }{9 } \frac{ M_Z^2 v^2 }{\Lambda^4 }  \frac{1}{\hat s^{3/2}} \frac{ 2- x^2  }{ x \sqrt{ 1- x^2}} \\
  \frac{\mathrm{d} \hat \sigma( q_i g \rightarrow \nu \bar \nu q_j)}{\mathrm{d} P_T } &\approx \frac{ \alpha_s { C^{ZP}_{i,j} }^2 \text{Br}\left( Z \rightarrow \nu \bar \nu\right)}{24  }  \frac{ M_Z^2 v^2}{ \Lambda^4 } \frac{1}{\hat s^{3/2}} \frac{ 4 +  x^2}{ x \sqrt{1-x^2}} 
\end{align}
Note that $C^{ZP}_{i,j} $ here is the definition from equation \eqref{eqn:WC_eff_penguin}, since at parton level the flavors do not mix and therefore the definition holds for all $i,j$.
\subsubsection{EW dipole operators}
The $P_T$-spectra are then given in the NWA
\begin{align}
  \frac{\mathrm{d} \hat \sigma( q_i \bar q_j \rightarrow \nu \bar \nu g)}{\mathrm{d} P_T } &= \frac{4 v^2 \alpha_s {C^{EW}_{i,j}}^2 \text{Br}\left( Z \rightarrow \nu \bar \nu\right) }{9  \Lambda^4   } \frac{ M_Z^2 \hat s ( \hat s - 4 P_T^2) + 4 P_T^2 \hat s^2+ M_Z^6}{ P_T \hat s^2  \sqrt{ ( \hat s- M_Z^2 )^2 - 4 P_T^2 \hat s} }    \\
  \frac{\mathrm{d} \hat \sigma( q_i g \rightarrow \nu \bar \nu q_j)}{\mathrm{d} P_T } &= \frac{- v^2 \alpha_s  {C^{EW}_{i,j}}^2  \text{Br}\left( Z \rightarrow \nu \bar \nu\right) }{12  \Lambda^4  } \\ \nonumber
  &\times \frac{ \left( 3 M_Z^4 \hat s ( P_T^2 - \hat s) + M_Z^2 \hat s^2 (P_T^2 + \hat s) + 4 P_T^2 \hat s^3 - 2 M_Z^8 + 4 M_Z^6 \hat s \right) }{ P_T \hat s^3 \sqrt{ (\hat s- M_Z^2)^2 - 4 P_T^2 \hat s} }
\end{align}
In the high energy limit these approach
\begin{align}
  \frac{\mathrm{d} \hat \sigma( q_i \bar q_j \rightarrow \nu \bar \nu g)}{\mathrm{d} P_T } &\approx  \frac{4 \alpha_s {C^{EW}_{i,j}}^2 \text{Br}\left( Z \rightarrow \nu \bar \nu\right) }{9  } \frac{v^2}{\Lambda^4}  \frac{1}{\sqrt{\hat s} }\frac{ x}{\sqrt{ 1 - x^2}} \\
  \frac{\mathrm{d} \hat \sigma( q_i g \rightarrow \nu \bar \nu q_j)}{\mathrm{d} P_T }     &\approx  \frac{ \alpha_s {C^{EW}_{i,j}}^2 \text{Br}\left( Z \rightarrow \nu \bar \nu\right) }{6   }  \frac{v^2}{\Lambda^4} \frac{1}{\sqrt{\hat s} } \frac{ x}{\sqrt{ 1- x^2}} 
\end{align}
\subsubsection{G dipole operators}
The $P_T$-spectra are then given in the NWA, for the up- and down-sector,
\begin{align}
   \frac{\mathrm{d} \hat \sigma( q_i \bar q_j \rightarrow \nu \bar \nu g)}{\mathrm{d} P_T } &= \frac{ {C^{G}_{i,j}}^2  \text{Br}\left( Z \rightarrow \nu \bar \nu\right) v^2}{9 \pi M_Z^2\Lambda^4 } \frac{P_T \left( 2 M_Z^2 \left({\epsilon_L^{i,j}}^2 + {\epsilon_R^{i,j}}^2\right)   + \left(\epsilon_L^{i,j}- {\epsilon_R^{i,j}}\right)^2  P_T^2 \right)}{ \sqrt{ (M_Z^2 - \hat s )^2 - 4 P_T^2 \hat s}} \\
   \frac{\mathrm{d} \hat \sigma( q_i g \rightarrow \nu \bar \nu q_j)}{\mathrm{d} P_T } &= \frac{ {C^{G}_{i,j}}^2  \text{Br}\left( Z \rightarrow \nu \bar \nu\right)  v^2 }{48 \pi M_Z^2\Lambda^4 } \\ \nonumber
   &\times \left(  \hat s - M_Z^2 \right) \frac{P_T \left( 2 M_Z^2 \left({\epsilon_L^{i,j}}^2 + {\epsilon_R^{i,j}}^2\right)   + \left(\epsilon_L^{i,j}- {\epsilon_R^{i,j}}\right)^2  \hat s^2 \right)}{ \hat s \sqrt{ (M_Z^2 - \hat s )^2 - 4 P_T^2 \hat s}} 
\end{align}
In the high energy limit the different sectors converge to the same result 
\begin{align}
  \frac{\mathrm{d} \hat \sigma( q_i \bar q_j \rightarrow \nu \bar \nu g)}{\mathrm{d} P_T } &\approx \frac{ {C^{G}_{i,j}}^2 \text{Br}\left( Z \rightarrow \nu \bar \nu\right) \left(\epsilon_L^{i,j}- {\epsilon_R^{i,j} }\right)^2 v^2}{72 \pi M_Z^2  } \frac{1}{\Lambda^4}\sqrt{\hat s} \frac{x^3 }{ \sqrt{ 1- x^2} } \\
  \frac{\mathrm{d} \hat \sigma( q_i g \rightarrow \nu \bar \nu q_j)}{\mathrm{d} P_T } &\approx \frac{ {C^{G}_{i,j}}^2  \text{Br}\left( Z \rightarrow \nu \bar \nu\right)\left(\epsilon_L^{i,j}- {\epsilon_R^{i,j} }\right)^2 v^2}{96 \pi M_Z^2  } \frac{1}{\Lambda^4} \sqrt{\hat s} \frac{x  }{\sqrt{ 1- x^2} }
\end{align}

\bibliography{cite}

\end{document}